\documentclass[%
 reprint,
 amsmath,amssymb,
 aps,
prb,
]{revtex4-1}
\usepackage{graphicx}
\usepackage{nicefrac}
\usepackage{amsfonts}
\usepackage{amssymb}
\usepackage{amsmath} 
\usepackage{subfigure}
\usepackage{multirow} 
\usepackage{tabularx}
\usepackage{array}
\usepackage{units}
\usepackage{tensor} 
\usepackage{braket}
\usepackage{multirow}
\usepackage{hyperref}

\usepackage{newtxtext,newtxmath}
\renewcommand{\bm}[1]{\boldsymbol{#1}}
\begin{document}
\preprint{APS/123-QED}
\title{First-principles study of LaOPbBiS$_3$\\and its analogous compounds as thermoelectric materials}
\author{Keiya Kurematsu$^{1)}$, Masayuki Ochi$^{1)}$, Hidetomo Usui$^{2)}$, and Kazuhiko Kuroki$^{1)}$}
\affiliation{$^{1)}$Department of Physics, Osaka University, Machikaneyama-cho, Toyonaka, Osaka 560-0043, Japan\\
			$^{2)}$Department of Material Science, Shimane University, Matsue, Shimane 690-8504, Japan}
\date{\today}

\begin{abstract}
LaOBiPbS$_3$ is a kind of pnictogen-dichalcogenide layered compounds, which have recently been experimentally investigated as thermoelectric materials owing to their low thermal conductivity and high controllability of constituent elements.
However, thermoelectric performance of LaOBiPbS$_3$ is at present not very high and that of its analogous compounds remains to be unknown.
In this study, we theoretically investigate thermoelectric properties of 24 possible variations of the constituent elements in LaOBiPbS$_3$ from the viewpoint of the electronic structure.
We find that some compounds can have much better thermoelectric performance than LaOBiPbS$_3$;
in particular, LaOSbPbSe$_3$ is predicted to have a power factor five times as large as that of LaOBiPbS$_3$.
Here, the choice of the pnictogen atom (As, Sb, and Bi), of which the low-energy conduction bands mainly consist, correlates with the calculated power factor and the dimensionless figure of merit, $ZT$.
Such correlation comes from the fact that the low-dimensionality of the electronic structure, which enhances the density of states near the band edge, strongly depends on the pnictogen atom through, e.g., the strength of the spin-orbit coupling.
Moreover, hybridization of the wave functions in the pnictogen-dichalcogenide layer and those in the rock-salt layer plays a key role in gap opening, and thus is important for achieving high thermoelectric performance.
In LaOSbPbSe$_3$, such hybridization also pushes up the conduction band bottom, which enhances the density of states near the band edge and thus the power factor.
\end{abstract}

\maketitle

\section{introduction}
Thermoelectric generation has recently been receiving a lot of attention because of its importance in energy harvesting, and many researchers seek high-performance thermoelectric materials for putting it to partial use.
The efficiency of thermoelectric conversion is assessed by the dimensionless figure of merit, $ZT$:
\begin{align}
ZT&=\frac{\sigma S^2}{\kappa}T=\frac{\text{PF}}{\kappa_{\mathrm{el}}+\kappa_{\mathrm{ph}}}T, \label{eq:ZT}
\end{align}
where $\sigma,\,S,\,\kappa$ and $T$ are the electrical conductivity, the Seebeck coefficient, the thermal conductivity, and the temperature, respectively.
In the rightmost term of Eq.~(\ref{eq:ZT}), the thermal conductivity is written as the sum of the electronic and the phononic contribution: $\kappa = \kappa_{\mathrm{el}}+\kappa_{\mathrm{ph}}$, and the power factor is defined as follows: $\text{PF}=\sigma S^2$, which is also used for measuring the thermoelectric performance.
Through a long attempt to increase $ZT$ and PF, several favorable aspects of the electronic structure were found, such as the multi-valley band structure~\cite{convergince} and the low-dimensionality~\cite{lowdim,pudding,PhysRevB.47.12727,cabon}.

From this viewpoint, pnictogen-dichalcogenide layered compounds~\cite{JPSJ.88.041001}, such as Bi$_4$O$_4$S$_3$ and $Ln$OBiS$_2$ ($Ln$ = La, Nd, Ce, etc.), which have been well-known as a superconductor~\cite{supercon,PhysRevB.86.220510,MIZUGUCHI201534,YAZICI2015218,supcomd.mat1.50.2015} and recently gathered attention also as a thermoelectric material~\cite{series}, are promising because they possess low-dimensional as well as multi-valley band structure, owing to the quasi-one-dimensional character of the $Pn$-$p_{x,y}$ orbitals in the conducting $PnCh_2$ layer ($Pn$ = Pnictogen = Bi, Sb; $Ch$ = Chalcogen = S, Se)~\cite{quasiband,LaOAsSe2,JPSJ.88.041010} alternately stacked with the insulating layer.
A rich variety of the insulating layer, such as $Ln$O in $Ln$OBiS$_2$, is also preferable because it offers high controllability of the crystal structure.
In addition, it was shown that the rattling motion of the Bi atom possibly reduces the thermal conductivity in LaOBiS$_{2-x}$Se$_x$ \cite{lattring,JPSJ.88.041009}.
As a matter of fact, LaOBiSSe was reported to exhibit a relatively high $ZT\sim0.36$ at 650 K with the low thermal conductivity $\kappa \sim 0.8$--$1.2$ W m$^{-1}$ K$^{-1}$~\cite{SSeZT}.
It is noteworthy that a recent theoretical study~\cite{LaOAsSe2} pointed out that the thermoelectric performance of $Ln$OBiS$_2$ can be sizably enhanced by using lighter and heavier elements for the pnictogen and chalcogen atoms, respectively.
Along this line, recent experimental findings are intriguing: $Ln$OSbSe$_2$ exhibits a low thermal conductivity, 1.5 W m$^{-1}$ K$^{-1}$ for $Ln$ = La and 0.8 W m$^{-1}$ K$^{-1}$ for $Ln$ = Ce~\cite{JPSJ.87.074703}, and $ZT$ is enhanced by a partial substitution of Sb for Bi in NdO$_{0.8}$F$_{0.2}$BiSe$_{2}$~\cite{JPSJ880247052019}.

\begin{figure}
	\centering
	\includegraphics[width=1\columnwidth]{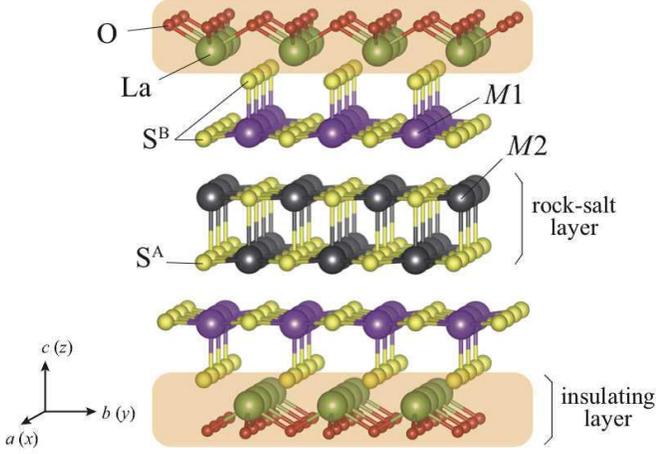}
	\caption{Crystal structure of LaO(PbS)BiS$_2$. In this study, we assumed that the $M$1 and $M$2 sites are occupied by Bi and Pb atoms, respectively (see the main text for detail). Depicted using the VESTA software~\cite{VESTA}.}
	\label{fig:1}
\end{figure}

Given this background, searching thermoelectric materials with the $PnCh_2$ layer is an attractive idea.
One possible candidate is LaOBiPbS$_3$~\cite{firstsynth,siteslect}, where the BiS$_2$ conducting layer is stacked not only with the LaO insulating layer but also with the PbS rock-salt layer, as shown in Fig.~\ref{fig:1}.
We note that, while Bi and Pb located at the $M$1 and $M$2 sites are not completely ordered, it was shown that the $M$1 ($M$2) site is mainly occupied by Bi (Pb) atom~\cite{siteslect}.
From now on, this composition is denoted by LaO(PbS)BiS$_2$, by which one can easily see the constituent elements of each layer.
It is advantageous that the electron carrier doping is possible by a partial replacement of oxygen with fluorine~\cite{fdope}.
While the thermoelectric performance of LaO(PbS)BiS$_2$ is at present not very high, e.g., PF$\,\sim1.5$$\mu$W cm$^{-1}$ K$^{-2}$ at 770 K~\cite{fdope},
a possible enhancement of its performance is expected through a change of the constituent elements because of its large degrees of freedom,
most of which still have remained unexplored.

In this study, we theoretically investigate the thermoelectric properties of LaO($TtCh^{\text{A}}$)$PnCh^{\text{B}}_{2}$ ($Pn$ = As, Sb, Bi; $Tt$ = Tetrel = Sn, Pb; $Ch^{\text{A,\,B}}$ = S, Se).
We find a correlation between the choice of the $Pn$ atom, of which the low-energy conduction bands mainly consist, and the calculated PF and $ZT$.
Such correlation comes from the fact that the low-dimensionality of the electronic structure, which enhances the density of states (DOS) near the band edge, strongly depends on the $Pn$ atom through, e.g., the strength of the spin-orbit coupling (SOC).
In addition, hybridization of the wave functions in the $PnCh^{\text{B}}_2$ layer and those in the $TtCh^{\text{A}}$ rock-salt layer plays a key role in gap opening, and thus is important for high thermoelectric performance.
A calculated power factor of LaO(PbS)BiS$_2$ is enhanced by a factor of five in LaO(PbSe)SbSe$_2$, where the hybridization mentioned above also pushes up the conduction band bottom, which enhances DOS near the band edge and thus the power factor.

This paper is organized as follows.
Section II presents the detailed procedure of our calculation and a brief description of the Boltzmann transport theory used in our work.
We present the basic thermoelectric properties of LaO(PbS)BiS$_2$ in Sec. III A.
Section III B presents calculated PF and $ZT$ for all the chemical compositions investigated in this study, and theoretical analysis of it to clarify the role of each atom in LaO($TtCh^{\text{A}}$)$PnCh^{\text{B}}_{2}$. Section IV is devoted to the conclusion of this study.

\section{calculation method}
First, we optimized the crystal structure of our target materials using the Perdew-Burke-Ernzerhof exchange-correlation functional revised for solids (PBEsol)~\cite{PBEsol} and the projector-augmented wave method~\cite{PhysRevB.59.1758} as implemented in the {\it Vienna ab initio simulation package} (VASP)~\cite{vasp1, vasp2, vasp3, vasp4}.
Our calculation started from the experimental crystal structure of LaO(PbS)BiS$_2$ taken from Ref.~\cite{siteslect}, and kept its symmetry, i.e., a space group of $P4/nmm$ and the Wyckoff positions of the atoms therein, during the structural optimization.
In addition, we assumed that our target material consists of the LaO, $TtCh^{\text{A}}$, and $PnCh^{\text{B}}_{2}$ layers, i.e., we neglected the antisite occupation of the $Tt$ and $Pn$ sites.
Further detailed analysis of the crystal symmetry and the antisite occupation is an important but challenging future issue.
In the structural optimization, we used a $10\times10\times3\,\bm{k}\,$mesh and a plane-wave cutoff energy of 550 eV with the inclusion of SOC.
Obtained lattice constants for all the materials investigated in this study are shown in Appendix~\ref{sec:app}.

Using the optimized crystal structure, we next performed band-structure calculation using the modified Becke-Johnson potential proposed by Tran and Blaha~\cite{mbj1, mbj2} and the full-potential (linearized) augmented plane-wave method as implemented in the \textsc{wien2k} package~\cite{wien2k}.
We used a $16\times16\times3\,\bm{k}\,$mesh and set the $RK_{\text{max}}$ parameter to 7. SOC was included unless noted.
Finally, we calculated the transport properties using the BoltzTraP package~\cite{BoltzTraP}.
Using this package, one can calculate the following quantities on the basis of the Boltzmann transport theory with the constant relaxation-time approximation,
\begin{align}
	\bm{\sigma}=e^2&\bm{K}_{0}\,,\hspace{2em}\bm{S}=-\frac{1}{eT}\bm{K}_{0}^{-1}\bm{K}_{1}\,,\label{eq:kappa}\\
	\bm{\kappa}_{\mathrm{el}}&=\frac{1}{T}[\bm{K}_2-\bm{K}_{1}\bm{K}_{0}^{-1}\bm{K}_{1}]\,,\label{eq:sigmaseebeck}
\end{align}
where $e(>0)$ is the elementary charge and the $\bm{K}_{\nu}$ is the transport coefficient expressed as follows:
\begin{align}
\bm{K}_{\nu}=\tau\sum_{n,\bm{k}}\bm{v}_{_{n,\bm{k}}}\otimes\bm{v}_{_{n,\bm{k}}}\left[-\frac{\partial f_{0}}{\partial E_{n,\bm{k}}}\right]\left(E_{n,\bm{k}}-\mu(T)\right)^{\nu},\label{eq:Ktensor}
\end{align}
where $\tau,\,\bm{v}_{_{n,\bm{k}}},\,f_{0},\,E_{n,\bm{k}}, \mu(T)$ are the relaxation time, the group velocity of the electron on the $n$-th band at the certain Bloch wave vector $\bm{k}$, the Fermi-Dirac distribution function, electronic energy dispersion and the chemical potential, respectively.
For calculating these quantities, we used a $62\times62\times12\,\bm{k}$\,mesh and the temperature $T=300\,\text{K}$.
Because an experimental study on LaO$_{1-x}$F$_{x}$(PbS)BiS$_2$ succeeded in performing electron-carrier doping by a partial replacement of oxygen with fluorine~\cite{fdope},
we concentrated on the transport properties in the electron-doped region using the rigid-band approximation.
In addition, we concentrated on the in-plane diagonal components of these tensors, e.g., $\text{PF}=\text{PF}_{xx}=\text{PF}_{yy}$, because the off-diagonal components vanish for the space group of $P4/nmm$ and the electronic conductivity along the $z$ axis is very small owing to the existence of the insulating layer.
Because it is challenging to determine the relaxation time $\tau$ and the lattice thermal conductivity $\bm{\kappa}_{\mathrm{ph}}$ with first-principles calculation, we assumed $\tau=5.0\times10^{-15}\,\text{s}$ and $\kappa_{\mathrm{ph}}=3.0\,\text{W\,m}^{-1}\,\text{K}^{-1}$ in this study.
This value of the relaxation time is typical for thermoelectric materials.
The lattice thermal conductivity here was determined by reference to the experimental total thermal conductivity of LaO(PbS)BiS$_2$, $\kappa \sim$ 1--4 $\text{W\,m}^{-1}\,\text{K}^{-1}$ at 50--300 K~\cite{firstsynth}.

To obtain the orbital-decomposed band dispersion, we extracted the Wannier functions from the first-principles band structure without the maximal localization procedure by using the \textsc{wien2wannier} and \textsc{wannier90} packages~\cite{MOSTOFI2008685,KUNES20101888,PhysRevB.65.035109,PhysRevB.56.12847}. We took the $p$ orbitals of all the atoms except for La as the Wannier functions, using a $10\times10\times3\,\bm{k}\,$mesh in the Wannier construction.
We also used these Wannier functions to see the role of the rock-salt layer in the thermoelectric performance, which we shall describe later in Sec.III B 4. In Sec. III B 4, we also calculated the transport properties using the tight-binding model consisting of the Wannier functions. For this purpose, we used a $240\times240\times60~\bm{k}$-mesh.

\section{results and discussion}
	\subsection{Thermoelectric properties of LaO(PbS)BiS$_2$\label{sec:result1}}
	
\begin{figure}
	\centering
	\includegraphics[width=1\columnwidth]{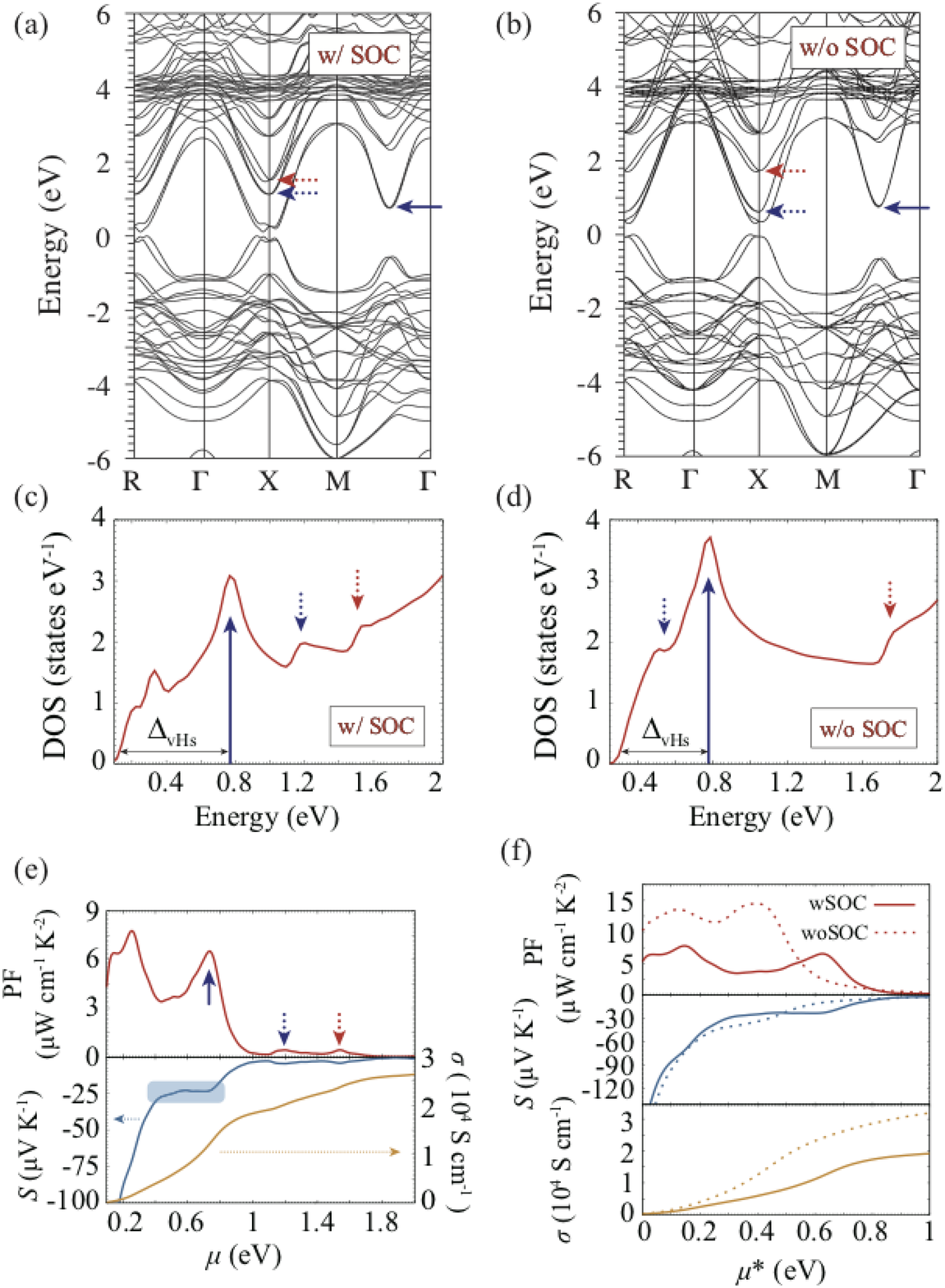}
	\caption{(a)--(b) First-principles band structure of LaO(PbS)BiS$_2$ calculated with and without SOC, respectively. The energy of the valence-band top was set to zero. (c)--(d) DOS of LaO(PbS)BiS$_2$ calculated with and without SOC, respectively. The definition of $\Delta_{\text{vHs}}$ is described in the main text. (e) PF, $S$, and $\sigma$ of LaO(PbS)BiS$_2$ calculated with SOC at $T=$ 300 K. (f) Comparison of the transport properties at $T=$ 300 K calculated with and without SOC. The definition of $\mu^*$ is described in the main text.}
	\label{fig:2}
	\end{figure}

	To begin with, we investigated the electronic structure of LaO(PbS)BiS$_2$.
	Figure~\ref{fig:2}(a) presents its first-principles band structure calculated with SOC.
	As shown by first-principles calculation presented in Ref.~\cite{siteslect}, the valence and conduction bands mainly consist of the Bi-$p$ orbitals in the BiS$_2$ layer and the S-$p$ orbitals in the PbS rock-salt layer, respectively.
	In other words, the band structure of LaO(PbS)BiS$_2$ near the valence-band top can be regarded as a consequence of the hybridization between the conduction bands of the BiS$_2$ layer and the valence bands of the PbS rock-salt layer.
	We note that the conduction bands here are quite similar to those in LaOBiS$_2$, which has a stacked structure of the LaO insulating layer and the BiS$_2$ conducting layer~\cite{quasiband}.
	A difference between them, except the hybridization mentioned above, is the existence of the Pb-$p$ band edge in LaO(PbS)BiS$_2$, which is indicated by a red dotted arrow (a dotted arrow with a higher energy at the X point) in Fig.~\ref{fig:2}(a).
	Another dotted arrow colored in blue (a dotted arrow with a lower energy at the X point) in Fig.~\ref{fig:2}(a) denotes the band edge of the Bi-$p$ bands. At the X = ($\pi a^{-1}$, 0, 0) point, the Bi-$p_{x,y}$ bands form the band edges with different energies, one of which lies at the energy shown with the blue (lower-energy) dotted arrow and the other of which lies at the valence-band top in Fig.~\ref{fig:2}(a).
	While the energy difference between these band edges are small when calculated without SOC as shown in Fig.~\ref{fig:2}(b), SOC enhances the hybridization between the Bi-$p_{x,y}$ orbitals, and then the energy difference increases as shown in Fig.~\ref{fig:2}(a), in the same manner as LaOBiS$_2$~\cite{LaOAsSe2}.
	
	Here, we point out that each band dispersion near the Fermi energy exhibits nearly two-fold degeneracy without the spin degrees of freedom, because there are two BiS$_2$ layers and two PbS planes in the unit cell of LaO(PbS)BiS$_2$ (see Fig.~\ref{fig:1}).
	Because these two BiS$_2$ layers are separated by the rock-salt layer, the band splitting caused by the bilayer coupling between the BiS$_2$ layers is almost absent for the Bi-$p$ bands. As for LaOBiS$_2$, where the BiS$_2$ layers are neighboring,
	such a band splitting is actually small, partially by the symmetry of the crystal structure and the Bloch wave functions~\cite{JPSJ.85.094705}, but larger than that in LaO(PbS)BiS$_2$.
	
	The last feature of the band structure we mention here is that there is a van Hove singularity (vHs) at the energy of around 0.8 eV indicated by blue solid arrows in Fig.~\ref{fig:2}(a), which is the same feature as LaOBiS$_2$~\cite{quasiband}. 
	We can verify the interesting aspects mentioned above also by looking into DOS shown in Figs.~\ref{fig:2}(c)--(d). Here, $\Delta_{\text{vHs}}$ is the energy difference between the vHs indicated by the solid arrows in Figs.~\ref{fig:2}(a)--(d) and the conduction band bottom. 
	
	Figure~\ref{fig:2}(e) presents the power factor PF, the Seebeck coefficient $S$, and the electrical conductivity $\sigma$, calculated at 300 K with the inclusion of SOC.
	We can find that the band edges and vHs indicated by the arrows in Fig.~\ref{fig:2}(c) result in peaks of PF.
	It is interesting that the PF peak around $\mu = 0.7$ eV has a comparable height to the PF peak near the conduction band bottom, even though the former is energetically far away from the conduction band bottom. As a matter of fact, the band edges at the X point with higher energies indicated by dotted arrows in Fig.~\ref{fig:2}(a) cannot yield a large PF peak as shown in Fig.~\ref{fig:2}(e), because the Seebeck coefficient becomes too small by such a heavy electron doping. The PF peak around $\mu=0.7$ eV owes to the large DOS of vHs, which prevents the Seebeck coefficient from decreasing by increasing the chemical potential, and also augments the electrical conductivity.
	If this vHs gets close to the conduction band bottom, i.e., $\Delta_{\text{vHs}}$ becomes small, the thermoelectric performance is expected to be much improved.
	As pointed out in Ref.~\cite{LaOAsSe2}, the reduction of $\Delta_{\text{vHs}}$ is closely related to the enhancement of the quasi-one-dimensionality of the electronic structure.
	In fact, the conduction band shown in Fig.~\ref{fig:2}(a) has a sharp dispersion along the $\Gamma$-X-M line, but a small $\Delta_{\text{vHs}}$ corresponds to a less-dispersive nature of the band dispersion along certain directions (concretely, from X to the center of the $\Gamma$-M line).
	Moreover, the insulating layer makes the band dispersion along the $k_z$ direction nearly flat.
	Therefore, the band dispersion hosts a quasi-one-dimensionality.
	In real space, such a quasi-one-dimensionality can be interpreted as a manifestation of the anisotropy of the in-plane Bi- and S-$p$ orbitals, which form the square lattice in the BiS$_2$ conducting layer, similarly to LaOBiS$_2$~\cite{quasiband}.
	It is advantageous for high thermoelectric performance that such a quasi-one-dimensional band dispersion is degenerate with respect to the spin degrees of freedom and nearly degenerate with respect to the bilayer degrees of freedom. In addition, the equivalency between the $x$ and $y$ directions for the crystal symmetry yields the multi-valley character: one quasi-one-dimensional band is dispersive with respect to the $(k_x+k_y)/\sqrt{2}$ direction and the other for the $(k_x-k_y)/\sqrt{2}$ direction, corresponding to the anisotropy of the $p_{(x\pm y)/\sqrt{2}}$ orbitals along the Bi-S bonds~\cite{quasiband}.
	
	It is noteworthy that decreasing $\Delta_{\text{vHs}}$ (i.e., enhancing the low-dimensionality) was theoretically verified to be a good strategy for enhancing the power factor in LaO$PnCh_2$~\cite{LaOAsSe2}.
	One strategy for reducing $\Delta_{\text{vHs}}$ in LaO$PnCh_2$ shown in that study is to decrease SOC, which seems to work well also for our system.
	Figure~\ref{fig:2}(f) presents the transport properties of LaO(PbS)BiS$_2$ calculated with and without SOC, where $\mu^*$ is defined as the chemical potential measured from the conduction band bottom.
	For metallic systems we shall investigate later in this paper, $\mu^*$ is the Fermi energy, which is set at zero, and so $\mu^*=\mu$ holds.
	As shown in Fig.~\ref{fig:2}(f), the electrical conductivity increases by turning off SOC, while keeping the Seebeck coefficient, which results in a sizable enhancement of PF. While the vHs is still far away from the conduction-band bottom even without SOC, an approach of the vHs to the band edge enhances DOS near the band edge as shown in Fig.~\ref{fig:2}(d).
	The change of the electronic structure here owes to the suppression of the splitting of the $Pn$-$p_{x,y}$ bands near the X point.
	Namely, the small splitting of the $Pn$-$p_{x,y}$ bands near the X point makes the energy difference between the vHs and the conduction band bottom, $\Delta_{\text{vHs}}$, smaller~\cite{LaOAsSe2}.
	We shall see later that the choice of the $Pn$ atom in LaO($TtCh^{\text{A}}$)$PnCh^{\text{B}}_{2}$ indeed affects its thermoelectric performance.
	
	We note that the present calculation result shows higher thermoelectric performance than those of experimental previous studies, e.g., PF$\,\sim1.5$ $\mu$W cm$^{-1}$ K$^{-2}$ at 770 K~\cite{fdope}, $\,\sim0.34$ $\mu$W cm$^{-1}$ K$^{-2}$ at 300 K~\cite{firstsynth}.
	In these experimental studies, the material has been synthesized as polycrystals, so that the thermoelectric efficiency is suppressed due to the presence of insulating layers in between the conducting layers. The present study indicates that the thermoelectric performance of this material can be enhanced if single crystals can be synthesized.

	\subsection{Elemental substitution}
	\begin{figure}
	\centering
	\includegraphics[width=1\columnwidth]{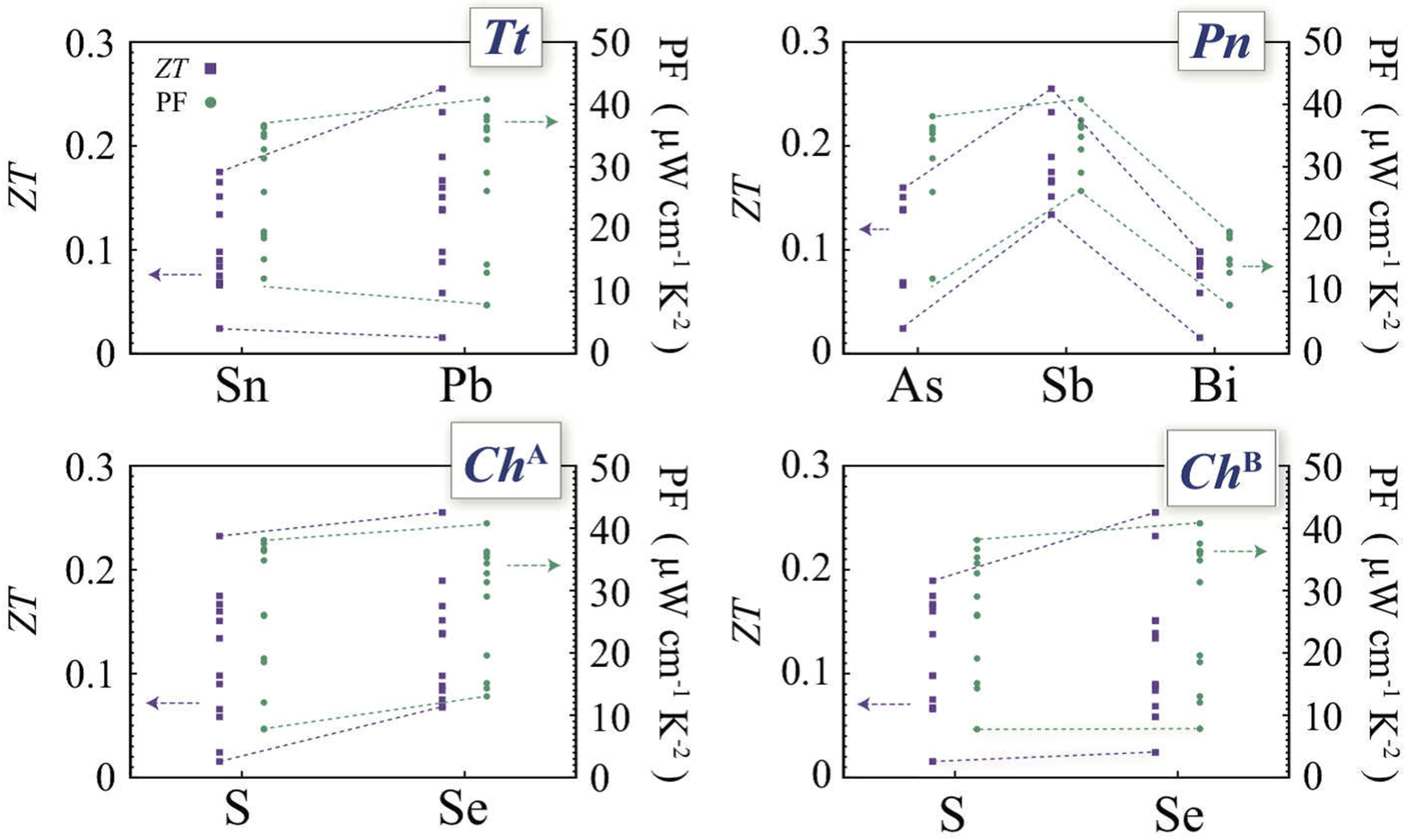}
	\caption{$ZT$ and PF values of LaO($TtCh^{\text{A}}$)$PnCh^{\text{B}}_{2}$ calculated at $T=$ 300 K using $\kappa_{\mathrm{ph}}=$ 3.0 W m$^{-1}$ K$^{-1}$ and $\tau = 5\times 10^{-14}$ s.
	In each plot, we classified 24 materials with $Tt$, $Pn$, $Ch^{\text{A}}$, and $Ch^{\text{B}}$ elements.
	The dashed lines connect the points with the lowest or highest $ZT$ or PF.}
	\label{fig:3}
	\end{figure}
	To seek the possibility of enhancing the thermoelectric performance of LaO(PbS)BiS$_2$, we have theoretically investigated 24 kinds of materials with the chemical composition LaO($TtCh^{\text{A}}$)$PnCh^{\text{B}}_{2}$ ($Pn$ = As, Sb, Bi; $Tt$ = Sn, Pb; $Ch^{\text{A,\,B}}$ = S, Se), which are analogous compounds of LaO(PbS)BiS$_2$.
	Using $\tau = 5\times 10^{-15}$ s and $\kappa_{\mathrm{ph}} =$ 3.0 W m$^{-1}$ K$^{-1}$, we calculated $ZT$ and PF for these 24 compositions as presented in Fig.~\ref{fig:3}.
	 In Fig.~\ref{fig:3}, we can find some trends for PF and $ZT$ with respect to constituent elements: for example, $ZT$ becomes higher for $Tt=$ Pb, $Pn=$ Sb, and $Ch^{\text{A, B}}=$ Se. In fact, we find that LaO(PbSe)SbSe$_2$ exhibits the highest thermoelectric performance in these candidate materials: PF $\sim 40$ $\mu$W cm$^{-1}$ K$^{-2}$ and $ZT$ $\sim 0.28$ for $T=300$ K.
	 From the next section, we shall see the microscopic origin of these trends in detail.
	 We note that these trends remain unaffected also when using $\tau=10^{-14}$ s, which we have checked since the relaxation time (to be more precise, the ratio $\tau \kappa_{\mathrm{ph}}^{-1}$) is regarded as an unknown parameter in this study.
		\subsubsection{Tt series}
		\begin{figure}[b]
			\centering
			\includegraphics[width=1\columnwidth]{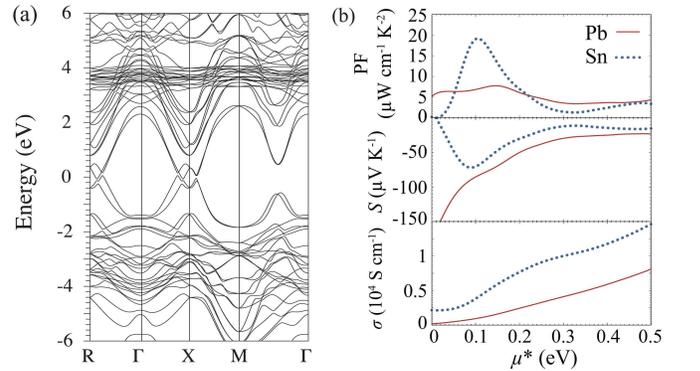}
			\caption{(a) Electronic band structure of LaO(SnS)BiS$_2$. (b) $\sigma$, $S$, PF, and (c) $ZT$ of LaO($Tt$S)BiS$_2$ for $Tt=$ Pb and Sn calculated at $T=$ 300 K.}
			\label{fig:4}
		\end{figure}
		To investigate the role of the $Tt$ element in the thermoelectric performance, we calculated the electronic band structure of LaO(SnS)BiS$_2$ as shown in Fig.~\ref{fig:4}(a). By comparing it with that of LaO(PbS)BiS$_2$ shown in Fig.~\ref{fig:2}(a), a crucial consequence by substituting Pb with Sn is the closing of the band gap.
		One possible cause of the gap closing is the weakened coupling between the $TtCh^{\text{A}}$ and $PnCh^{\text{B}}_2$ layers for $Tt=$ Sn.
		In fact, the $Pn$-$Ch^{\text{A}}$ distance becomes longer by substituting Pb with Sn as listed on Table~\ref{tab:1} in Appendix~\ref{sec:app}.
		Because the downward and upward convex bands around the Fermi energy mainly consist of the $Pn$- and $Ch^{\text{A}}$-$p$ orbitals, respectively~\cite{siteslect}, a long $Pn$-$Ch^{\text{A}}$ distance for $Tt=$ Sn will result in a smaller hybridization between these bands around the X point, and hence yield the metallic band structure. We note that all the compositions with $Tt=$ Sn investigated in this study exhibit a metallic band structure in our calculation, while some compositions with $Tt=$ Pb have a gapped one, as shown in Table~\ref{tab:1}.
		
		Fig.~\ref{fig:4}(b) present $\sigma$, $S$ and PF of LaO($Tt$S)BiS$_2$ with $Tt=$ Sn and Pb.
		At $\mu^*\simeq 0$, the metallic band structure for $Tt=$ Sn yields a small Seebeck coefficient, which drastically degrades PF.
		Although PF of $Tt=$ Sn becomes much higher than $Tt=$ Pb at $\mu^* \simeq 0.1$ eV because of the high electrical conductivity for the former, the merit in $ZT$ is limited because the high electrical conductivity inevitably results in a high electronic thermal conductivity ($\kappa_{\mathrm{el}}$), which appears in the denominator of $ZT$.
		While LaO(SnS)BiS$_2$ exhibits a maximum $ZT$ value larger than that of LaO(PbS)BiS$_2$, most of the $Tt=$ Sn group materials have smaller $ZT$ than the materials in the $Tt=$ Pb group, as shown in Table~\ref{tab:1}.
		The origin of this trend can be understood by the metallic band structure for $Tt=$ Sn.
		\subsubsection{Pn series}
		
		As we have seen in Fig.~\ref{fig:3}, a remarkable correlation exists between the $Pn$ element and the thermoelectric performance.
		Based on our observation described in the following analysis, there are two important ingredients that yield this correlation: quasi-one-dimensionality becomes strong by replacing Bi with lighter pnictogen atoms, but the arsenic makes the band structure metallic, which degrades, especially, $ZT$.
		
\begin{figure}
			\centering
			\includegraphics[width=1\columnwidth]{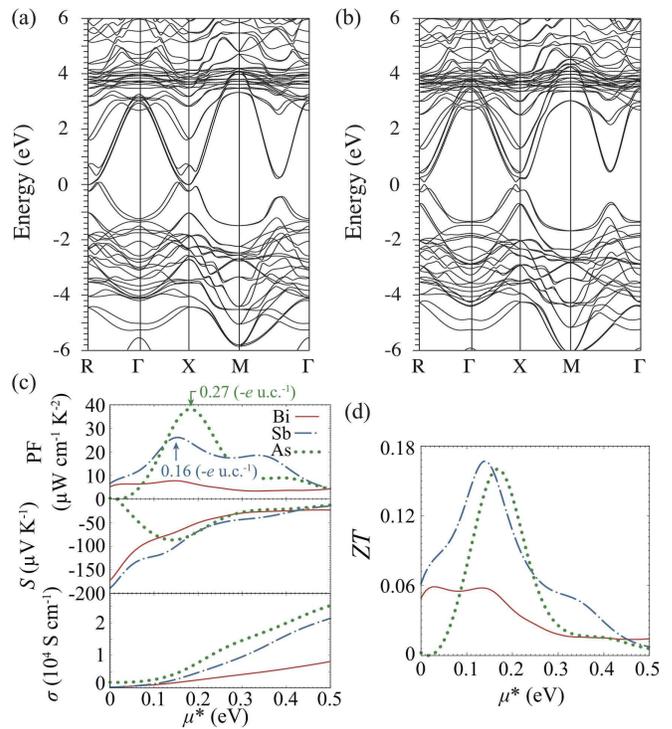}
			\caption{Electronic band structure of (a) LaO(PbS)AsS$_2$ and (b) LaO(PbS)SbS$_2$. (c) $\sigma$, $S$, PF, and (d) $ZT$ of LaO(PbS)$Pn$S$_2$ for $Pn=$ As, Sb, and Bi calculated at $T=$ 300 K. In panel (c), we show the carrier number per unit cell at the chemical potential where PF is maximized for the case of As and Sb.}
			\label{fig:5}
		\end{figure}

		Figures~\ref{fig:5}(a)--(b) present the electronic band structure of LaO(PbS)AsS$_2$ and LaO(PbS)SbS$_2$, respectively.
		The overall characteristics of the band structure are common to LaO(PbS)$Pn$S$_2$ ($Pn =$ As, Sb, Bi), but one important difference is the gap closing for $Pn=$ As.
		As a result, the transport properties, $\sigma$, $S$, and PF, of LaO(PbS)AsS$_2$ shown in Fig.~\ref{fig:5}(c) exhibit a behavior similar to those of LaO(SnS)BiS$_2$, which we have seen in the previous section.
		Namely, $S$ and PF become small at $\mu^* \simeq 0$ for LaO(PbS)AsS$_2$ because of its metallic band structure.
		Although PF becomes large by heavy carrier doping there, $ZT$ peak is smaller than that of $Pn=$ Sb, as shown in Fig.~\ref{fig:5}(d), which is likely be due to the large electronic thermal conductivity for $Pn=$ As with the metallic electronic structure.
		We note that all the other compositions with $Pn=$ As also have a metallic band structure as shown in Table~\ref{tab:1}.
		
		For $Pn=$ Sb, where the gap closing does not take place, PF and $ZT$ are much larger than those for $Pn=$ Bi as shown in Figs.~\ref{fig:5}(c)--(d).
		For LaOBiS$_2$, some of the present authors pointed out that a quasi-one-dimensionality of the electronic structure can be drastically enhanced by replacing Bi with Sb or As, because of the suppressed SOC, smaller inter-pnictogen transfer integrals, and the closer energy levels of the $Pn$ and S atomic orbitals~\cite{LaOAsSe2}.
		In Sec.~\ref{sec:result1}, we have already seen the effect of SOC on thermoelectric performance through the change in $\Delta_{\text{vHs}}$, i.e., the quasi-one-dimensionality, for our target material LaO(PbS)BiS$_2$.
		Because both LaO(PbS)BiS$_2$ and LaOBiS$_2$ have the BiS$_2$ conducting layer, the same strategy for improving the thermoelectric performance is expected to work well. As a matter of fact, we found that $\Delta_{\text{vHs}}$ decreases from 0.65 eV for LaO(PbS)(BiS$_2$) to 0.36 eV for LaO(PbS)(SbS$_2$).
		As a result, PF and $ZT$ are enhanced for $Pn=$ As and Sb compared with $Pn=$ Bi.
		We note that the enhancement in PF shown in Fig.~\ref{fig:5}(c) is larger than that induced by switching off SOC in LaO(PbS)BiS$_2$ shown in Fig.~\ref{fig:2}(f). This means that, while the suppression of SOC in $Pn=$ As or Sb plays some role in enhancing PF, other effects described above should also be important in our material. In addition, an important difference between LaO(PbS)$Pn$S$_2$ and LaO$Pn$S$_2$ is the possibility of the gap closing for the former, depending on the coupling between the $Pn$S$_2$ and PbS layers. This difference results in the metallic band structure and thus the degraded thermoelectric performance for $Pn=$ As as mentioned in the previous paragraph.
		\subsubsection{Ch series}
		\begin{figure}
			\centering
			\includegraphics[width=1.12\columnwidth]{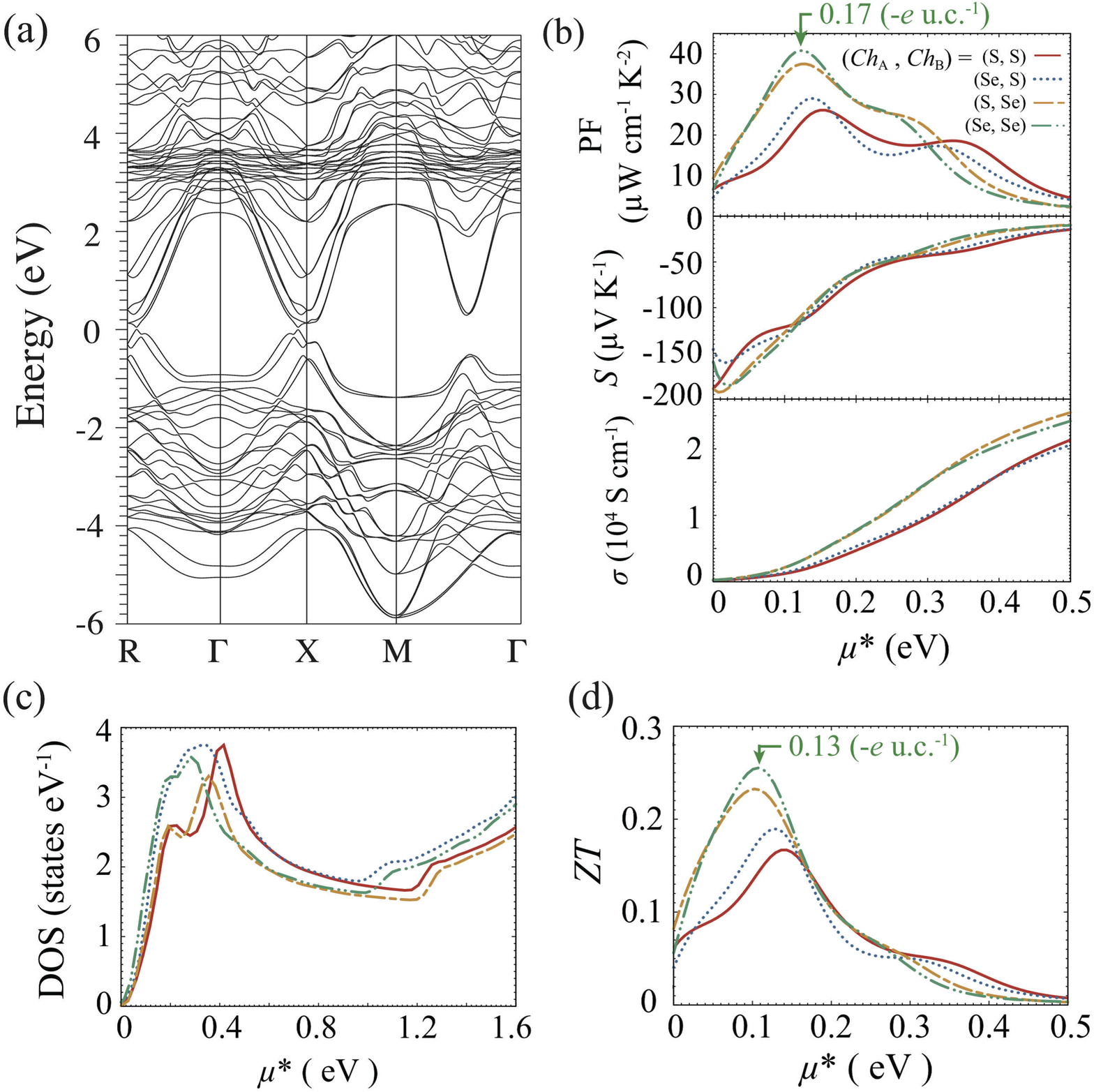}			
			\caption{(a) Electronic band structure of LaO(PbSe)SbSe$_2$. (b) $\sigma$, $S$, PF, (c) DOS, and (d) $ZT$ for LaO(Pb$Ch^{\text{A}}$)Sb$Ch^{\text{B}}_2$ ($Ch^{\text{A,B}} =$ S, Se). Calculation results shown in panels (b) and (d) were obtained using $T=$ 300 K. The carrier number per unit cell at the chemical potential where PF or $ZT$ are maximized for LaO(PbSe)SbSe$_2$ are shown in panels (b) and (d).}
			\label{fig:6}
		\end{figure}
		\begin{figure}[t]
			\centering
			\includegraphics[width=1\columnwidth]{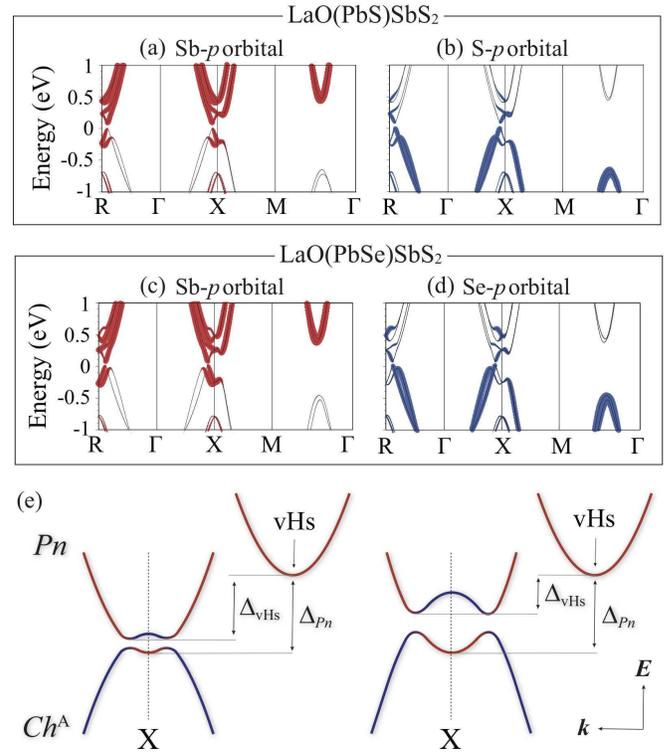}
			\caption{(a)--(b) The electronic band structure of LaO(PbS)SbS$_2$. The colored thickness of the lines indicates the weight of Sb (S$_{\text{rock-salt}}$)-$p$ orbital character. (c)--(d) The same plot for LaO(PbSe)SbS$_2$. (e) Schematic figure presenting the hybridization between $Pn$ and $Ch^{\text{A}}$.}
			\label{fig:7}
		\end{figure}
		Figure~\ref{fig:6}(a) presents the electronic band structure of LaO(PbSe)SbSe$_2$.
		By comparing it with the electronic band structure of LaO(PbS)SbS$_2$ shown in Fig.~\ref{fig:5}(b), we can find that the electronic band structure is less affected by the choice of the $Ch$ atom, except that the valence band structure is slightly shifted upward in Fig.~\ref{fig:6}(a) because the energy levels of the Se atomic orbitals are higher than those of the S atomic orbitals.
		
		However, by changing $Ch^{\text{B}}$ from S to Se, PF and $ZT$ are sizably enhanced.
		The calculated values of $\sigma$, $S$, PF, DOS, and $ZT$ for LaO(Pb$Ch^{\text{A}}$)Sb$Ch^{\text{B}}_2$ ($Ch^{\text{A,B}} =$ S, Se) are shown in Figs.~\ref{fig:6}(b)--(d). These quantities except of DOS were calculated at $T=$ 300 K.
		By looking into Fig.~\ref{fig:6}(c), DOS is less affected by $Ch^{\text{B}}$, but the electrical conductivity shown in Fig.~\ref{fig:6}(b) exhibits a sizable increase by the change of $Ch^{\text{B}}=$ S $\to$ Se, which means that the group velocity is enhanced by this change.
		A possible reason for it is the approaching energy levels of the Sb-$p$ orbitals and $Ch^{\text{B}}$-$p$ orbitals as discussed in Ref.~\cite{LaOAsSe2} for LaO$PnCh_2$, which strengthens the coupling between these atomic orbitals and then sharpens the band dispersion.
		Consequently, PF reaches a maximum value of nearly 40 $\mu$W cm$^{-1}$ K$^{-2}$. This value is similar to that of Bi$_2$Te$_3$, which is well known as a good thermoelectric material at room temperature. Although this comparison is based on a certain assumption for the $\tau$ value for the present material, the assumed value is still smaller than the known value for Bi$_2$Te$_3$. 
		
		On the other hand, the change of $Ch^{\text{A}}=$ S $\to$ Se only brings a small enhancement of PF and $ZT$ as shown in Figs.~\ref{fig:6}(b) and (d). This weak enhancement may come from the increased DOS as shown in Fig.~\ref{fig:6}(c).
		To investigate the origin of the DOS enhancement, we employed the Wannier orbitals to show the orbital character onto the band dispersion as presented in Figs.~\ref{fig:7}(a)--(d).
		We find that $\Delta_{\text{vHs}}$ becomes small by the change of $Ch^{\text{A}}=$ S $\to$ Se, from $0.30$ to $0.37$ eV.
		The situation is schematically shown in Fig.~\ref{fig:7}(e).
		Because of the different energy levels for the S- and Se-$p$ orbitals, the $Pn$ band bottom becomes a little far away from the conduction band bottom in $Ch^{\text{A}}=$ Se, as a consequence of the band hybridization shown here.
		Thus, $\Delta_{\text{vHs}}$, which is the energy difference between the conduction band bottom and vHs, becomes smaller in $Ch^{\text{A}}=$ Se.
		This means the quasi-one-dimensionality is enhanced, which enhances DOS near the band edge.
		
		For all the pairs of ($Ch^{\text{A}}$, $Ch^{\text{B}}$) for LaO(Pb$Ch^{\text{A}}$)Sb$Ch^{\text{B}}_2$, we verified that the band gap remains opened (see Table~\ref{tab:1}) and the characteristic band deformation as shown in Fig.~\ref{fig:7}(e) takes place at the X point. Such a deformed band dispersion reminds us of the pudding-mold-shaped band structure~\cite{JPSJ.76.083707}, which improves PF through the coexisting large DOS and the large group velocity for several materials~\cite{JPSJ.76.083707,Usui_2009,PhysRevB.88.075140,PhysRevB.88.075141,Usui2014}.
		
		We note that the effect of the chalcogen atoms on the thermoelectric performance depends on the $Tt$ and $Pn$ elements as shown in Table~\ref{tab:1}. In fact, the correlation between the thermoelectric performance, PF and $ZT$, and $Ch$ elements is not so strong as shown in Fig.~\ref{fig:3}. However, this weak correlation might be natural because of the relatively small difference of the band structure near the Fermi level induced by the chalcogen atoms, as we have seen in this section.
		\subsubsection{The effect of inserting rock-salt layer}
		\begin{figure}[b]
			\centering
			\includegraphics[width=1\columnwidth]{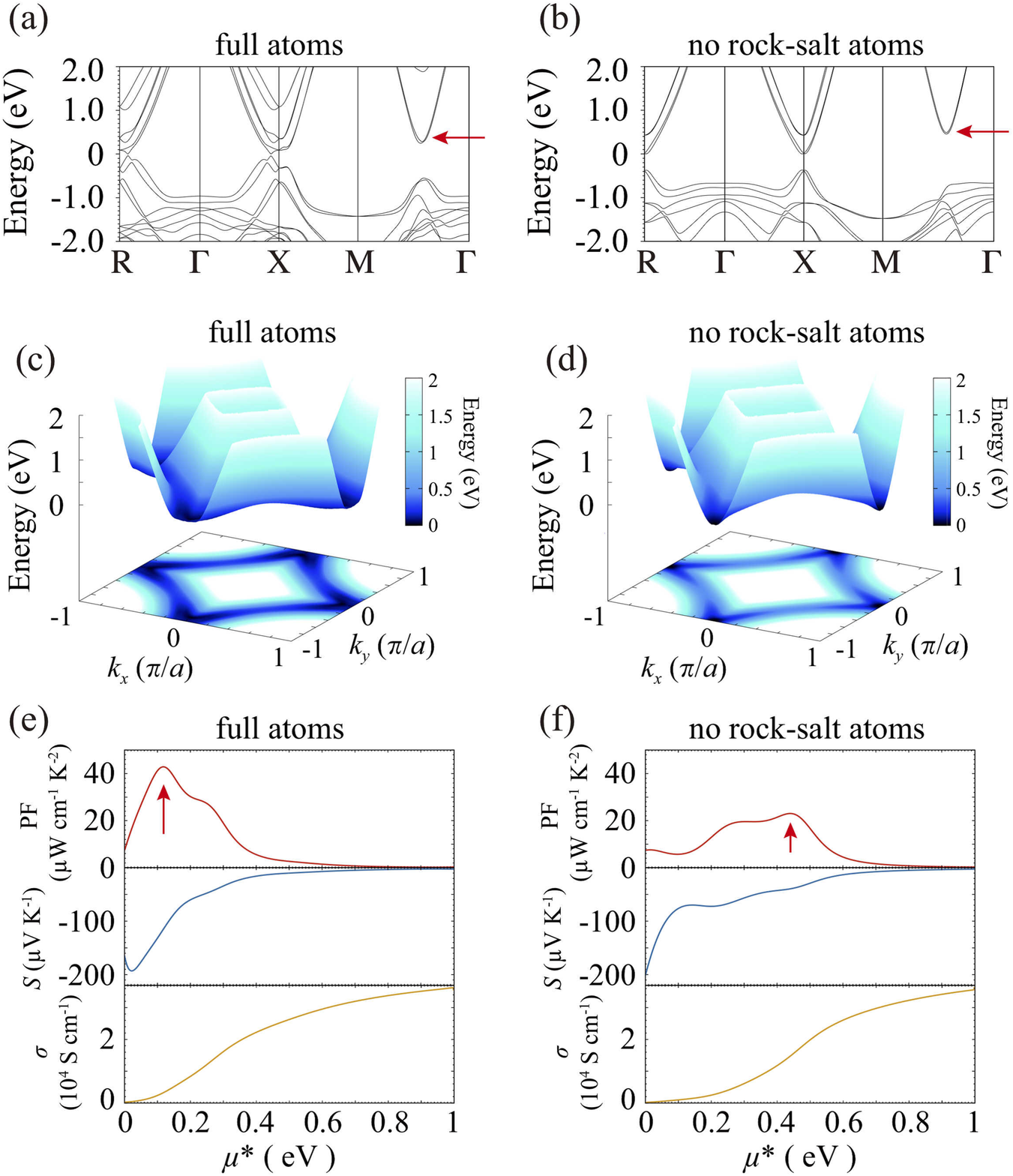}
			\caption{(a)--(b) Electronic band structures of two types of tight-binding model for LaO(PbSe)SbSe$_2$ and (c)--(d) those on the $k_z=0$ plane colored by the energy value. In panels (a)--(d), the energy of the conduction band bottom was set to zero. (e)--(f) Thermoelectric properties for these models.}
			\label{fig:8}
		\end{figure}
 		In this section, we focus on the effect of inserting rock-salt layer on thermoelectric performance.
		For this purpose, we calculated the electronic band structure and the thermoelectric properties by using two types of tight-binding models: one consists of all Wannier functions and the other consists of Wannier functions except for those in the rock-salt layer (the $Tt$-$p$ and the $Ch^{\text{A}}$-$p$ orbitals). 
 		Figures \ref{fig:8} (a)--(b) present the electronic band structures of two types of tight-binding model for LaO(PbSe)SbSe$_2$, and those on the $k_z=0$ plane are shown in Figs. \ref{fig:8} (c)--(d).
		In Fig. \ref{fig:8} (b), band dispersions originating from the rock-salt layer are not present because we used the tight-binding model not including Wannier functions in the rock-salt layer.
		On the other hand, in Figs. \ref{fig:8} (a), the band edge near the X point becomes less dispersive and $\Delta_{\text{vHs}}$ decreases because of the hybridization between the $Pn$-$p$ and the $Ch^{\text{A}}$-$p$ bands as discussed in Sec.III A.
		As shown in Figs. \ref{fig:8} (c)--(d), the band edge near the X point becomes like pudding-mold-shaped band structure (flattened band bottom) and the band dispersion along the lines $|k_x| + |k_y| = \pi$ becomes less dispersive (i.e. quasi-one-dimensionality is enhanced) by inserting the rock-salt layer.
		The comparison of the thermoelectric properties between two types of tight-binding models is presented in Figs.\ref{fig:8} (e)--(f) for LaO(PbSe)SbSe$_2$.
		In the case of LaO(PbSe)SbSe$_2$, $\Delta_{\text{vHs}}$ changes from 0.46~eV to 0.25~eV by inserting rock-salt layer, which results in a PF peak shift as shown with arrows in Figs. \ref{fig:8} (e)--(f).
		As a consequence of this PF peak shift, the maximum value of PF peak sizably increases by realizing both high Seebeck coefficient and high electrical conductivity.
		By inserting the rock-salt layer, the proportion of the volume of the conduction layer to the entire bulk is reduced, so that a reduction in the thermoelectric performance should be originally expected.
		However, it is surprising that there is instead an enhancement of PF.
		On the other hand, in the case of LaO(PbS)BiS$_2$ (shown in appendix B), although $\Delta_{\text{vHs}}$ becomes small by intercalating rock-salt layer (from 0.81 eV to 0.65 eV), vHs is still far away from conduction band bottom because of the effect of $Pn$, which we discussed in the Sec.III B 2.
		Seebeck coefficient becomes too small by such a heavy electron doping and hence PF is also suppressed.

		\subsubsection{Temperature dependence of $ZT$}	
		Because of the narrow (or even zero) band gap in our target materials, the temperature dependence of the thermoelectric performance is worth investigating.
		Figure 9 presents the temperature dependence of $ZT$ for LaO(PbSe)SbSe$_2$, which exhibits the highest $ZT$ within our calculation shown in this study.
		Here, the chemical potential is chosen at each temperature so as to realize the maximum $ZT$ at that temperature.
		We have tried three values for the relaxation time: $\tau=$ 1, 5, 10 fs here, which is expected to cover a typical range of the relaxation time in thermoelectric materials.
		\begin{figure}[b]
			\centering
			\includegraphics[width=0.9\columnwidth]{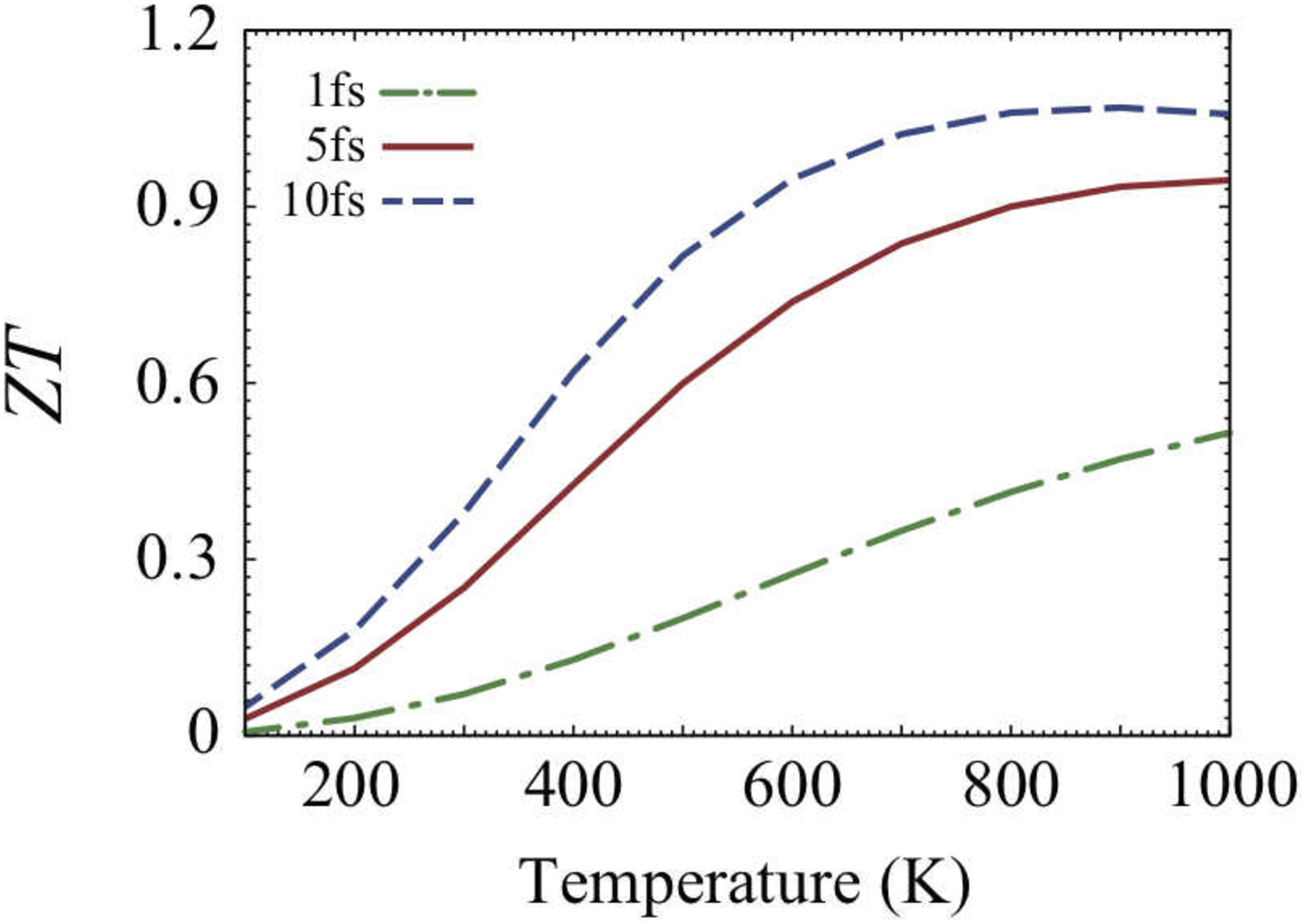}
			\caption{Temperature dependence of $ZT$ for LaO(PbSe)SbSe$_2$ calculated using several values of the relaxation time $\tau$. Here, the chemical potential is chosen at each temperature so as to realize the maximum $ZT$ at that temperature.}
			\label{fig:9}
		\end{figure}
		Normally, narrow-gap semiconductor, like this material, cannot exhibit high thermoelectric performance at high temperature, because both electrons and holes are excited, which reduces the Seebeck coefficient (bipolar effect).
		However, we found that high thermoelectric performance is realized at relatively high temperature in LaO(PbSe)SbSe$_2$:
		$ZT$ reaches around 0.9 at 1000 K for $\tau=5$ fs, and $1.1$ at 900 K for $\tau=10$ fs.
		Such a large $ZT$ owes to the deep chemical potential $\mu^*$ at the $ZT$ peak as shown in Fig.~\ref{fig:6}(d), which prevents the generation of the hole carriers.
		We note that, in reality, the heavy carrier doping and high temperature inevitably shorten the relaxation time through active electron-phonon scattering, which will somewhat suppress $ZT$ at deep $\mu^*$ and high temperature. Since the first-principles treatment of the electron-phonon scattering is challenging, this point is an important future issue.

\section{conclusion}
		We have theoretically investigated the thermoelectric performance (PF and $ZT$) of LaO($TtCh^{\text{A}}$)$PnCh^{\text{B}}_{2}$, and found that LaO(PbSe)SbSe$_2$ achieves the best performance among these 24 candidate materials.
		For LaO(PbSe)SbSe$_2$, $ZT$ reaches around 0.9 at 1000 K for $\tau=5$ fs, and $1.1$ at 900 K for $\tau=10$ fs.
		This high performance originates from the quasi-one-dimensionality of the electronic structure, which can be enhanced mainly by changing the $Pn$ atom, in the same manner as LaOBiS$_2$.
		The non-zero band gap is also an important factor for high thermoelectric performance. For example, $Tt=$ Sn is not favorable for achieving high $ZT$ because of the gap closing, induced by the suppressed coupling between the $TtCh^{\text{A}}$ rock-salt layer and the $PnCh^{\text{B}}_{2}$ conducting layer.
		Hybridization between the $PnCh^{\text{A}}$ and $TtCh^{\text{B}}_2$ layers is important not only for opening the gap, but also for making the van Hove singularity closer to the band edge, i.e., for enhancing the low-dimensionality and thus DOS near the band edge.
		Our study offers a possible designing principle for improving the thermoelectric performance of LaO(PbS)BiS$_2$.
\section*{acknowlegments}
We thank Yosuke Goto and Yoshikazu Mizuguchi for fruitful discussion.
This work was supported by JST-CREST Grant Number JPMJCR16Q6 and JSPS KAKENHI Grant Numbers JP17H05481, JP17K14108, JP18K13470, JP19H04697.
\appendix
\newcolumntype{Y}{>{\centering\arraybackslash}p{0.7cm}} 
\newcolumntype{Z}{>{\centering\arraybackslash}p{1.8cm}}
\newcolumntype{V}{>{\centering\arraybackslash}p{2.6cm}}
\newcolumntype{W}{>{\centering\arraybackslash}p{2cm}}
\section{The effect of inserting rock-salt layer}
Electronic band structures and thermoelectric properties of two types of tight-binding models for LaO(PbS)BiS$_2$ are shown in Fig. 10.
\begin{figure}
			\centering
			\includegraphics[width=1\columnwidth]{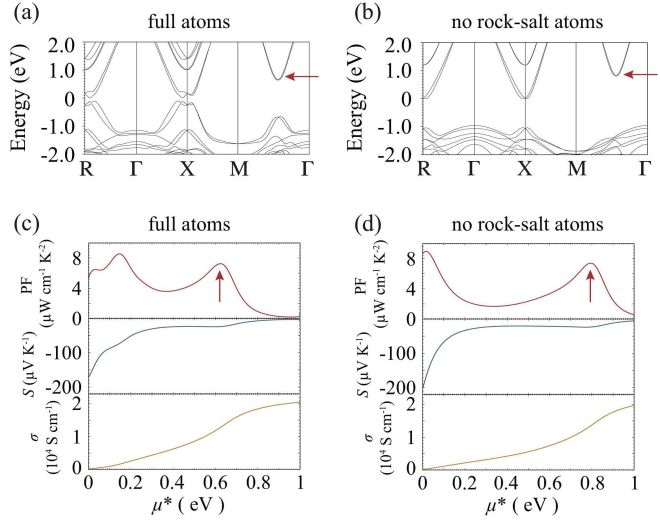}
			\caption{(a)--(b) Electronic band structures and (c)--(d) thermoelectric properties of two types of tight-binding models for LaO(PbS)BiS$_2$. In panels (a)--(b), the energy of the conduction band bottom was set to zero.}
			\label{fig:10}
		\end{figure}
		
\section{List of the optimized structural parameters, $ZT$, PF, and the band gap~\label{sec:app}}
Calculated values of the optimized structural parameters, the maximum $ZT$, the maximum PF, and the band gap of all the materials investigated in this study is shown in Table I.
\begin{table*}
			\caption{Calculated values of the optimized structural parameters, the maximum $ZT$, the maximum PF, and the band gap of all the materials investigated in this study. the maximum $ZT$ and the maximum PF were calculated at 300 K using $\tau=5.0\times10^{-15}\,\text{s}$ and $\kappa_{\mathrm{ph}}=3.0\,\text{W\,m}^{-1}\,\text{K}^{-1}$.}
			\centering
\begin{tabular}{Y|Y|Y|Y|ZZZ|ZV|W}
\hline\hline
$Tt$ & $Pn$ & $Ch^{\text{A}}$ & $Ch^{\text{B}}$ & $a$ (\AA) & $c$ (\AA) & $Pn$-$Ch^{\text{A}}$ (\AA) & $ZT$ & PF ($\mu$W cm$^{-1}$ K$^{-2}$) & Band gap (eV) \\\hline
\multirow{12}{*}{Pb} & \multirow{4}{*}{Bi} & S & S & 4.06 & 19.6 & 3.19 & 0.06 & 7.8 & 0.11 \\
 & & S & Se & 4.13 & 20.2 & 3.24 & 0.06 & 7.8 & 0.05 \\
 & & Se & S & 4.11 & 19.7 & 3.25 & 0.10 & 14.3 & 0 \\
 & & Se & Se & 4.18 & 20.3 & 3.29 & 0.09 & 13.0 & 0 \\
\cline{2-10}
 & \multirow{4}{*}{Sb} & S & S & 4.03 & 19.3 & 3.10 & 0.17 & 26.1 & 0.09 \\
 & & S & Se & 4.09 & 19.8 & 3.10 & 0.23 & 37.5 & 0.06 \\
 & & Se & S & 4.08 & 19.4 & 3.17 & 0.19 & 29.0 & 0.06 \\
 & & Se & Se & 4.14 & 20.0 & 3.17 & 0.26 & 40.8 & 0.04 \\
 \cline{2-10}
 & \multirow{4}{*}{As} & S & S & 3.97 & 18.7 & 2.99 & 0.16 & 38.1 & 0 \\
 & & S & Se & 4.02 & 19.3 & 2.97 & 0.15 & 36.3 & 0 \\
 & & Se & S & 4.02 & 18.8 & 3.12 & 0.14 & 34.3 & 0 \\
 & & Se & Se & 4.07 & 19.5 & 3.08 & 0.14 & 35.8 & 0 \\
\hline
\multirow{12}{*}{Sn} & \multirow{4}{*}{Bi}& S & S & 4.02 & 19.6 & 3.24 & 0.10 & 19.1 & 0 \\
 & & S & Se & 4.09 & 20.5 & 3.48 & 0.09 & 18.5 & 0 \\
 & & Se & S & 4.07 & 19.7 & 3.25 & 0.08 & 15.1 & 0 \\
 & & Se & Se & 4.14 & 20.4 & 3.35 & 0.08 & 19.6 & 0 \\
 \cline{2-10}
 & \multirow{4}{*}{Sb} & S & S & 3.99 & 19.2 & 3.14 & 0.17 & 36.7 & 0 \\
 & & S & Se & 4.06 & 19.9 & 3.21 & 0.13 & 34.8 & 0 \\
 & & Se & S & 4.04 & 19.3 & 3.16 & 0.17 & 32.7 & 0 \\
 & & Se & Se & 4.10 & 20.0 & 3.20 & 0.15 & 36.2 & 0 \\
 \cline{2-10}
 &\multirow{4}{*}{As} & S & S & 3.91 & 18.7 & 2.98 & 0.07 & 25.9 & 0 \\
 & & S & Se & 3.97 & 19.3 & 3.03 & 0.02 & 12.0 & 0 \\
 & & Se & S & 3.96 & 18.9 & 3.10 & 0.07 & 35.3 & 0 \\
 & & Se & Se & 4.02 & 19.5 & 3.11 & 0.07 & 31.3 & 0 \\ \hline\hline
\end{tabular}
\label{tab:1}
\end{table*}
\bibliography{LaOPbBiS3}

\providecommand{\noopsort}[1]{}\providecommand{\singleletter}[1]{#1}%
\begin{thebibliography}{44}%
\makeatletter
\providecommand \@ifxundefined [1]{%
 \@ifx{#1\undefined}
}%
\providecommand \@ifnum [1]{%
 \ifnum #1\expandafter \@firstoftwo
 \else \expandafter \@secondoftwo
 \fi
}%
\providecommand \@ifx [1]{%
 \ifx #1\expandafter \@firstoftwo
 \else \expandafter \@secondoftwo
 \fi
}%
\providecommand \natexlab [1]{#1}%
\providecommand \enquote  [1]{``#1''}%
\providecommand \bibnamefont  [1]{#1}%
\providecommand \bibfnamefont [1]{#1}%
\providecommand \citenamefont [1]{#1}%
\providecommand \href@noop [0]{\@secondoftwo}%
\providecommand \href [0]{\begingroup \@sanitize@url \@href}%
\providecommand \@href[1]{\@@startlink{#1}\@@href}%
\providecommand \@@href[1]{\endgroup#1\@@endlink}%
\providecommand \@sanitize@url [0]{\catcode `\\12\catcode `\$12\catcode
  `\&12\catcode `\#12\catcode `\^12\catcode `\_12\catcode `\%12\relax}%
\providecommand \@@startlink[1]{}%
\providecommand \@@endlink[0]{}%
\providecommand \url  [0]{\begingroup\@sanitize@url \@url }%
\providecommand \@url [1]{\endgroup\@href {#1}{\urlprefix }}%
\providecommand \urlprefix  [0]{URL }%
\providecommand \Eprint [0]{\href }%
\providecommand \doibase [0]{http://dx.doi.org/}%
\providecommand \selectlanguage [0]{\@gobble}%
\providecommand \bibinfo  [0]{\@secondoftwo}%
\providecommand \bibfield  [0]{\@secondoftwo}%
\providecommand \translation [1]{[#1]}%
\providecommand \BibitemOpen [0]{}%
\providecommand \bibitemStop [0]{}%
\providecommand \bibitemNoStop [0]{.\EOS\space}%
\providecommand \EOS [0]{\spacefactor3000\relax}%
\providecommand \BibitemShut  [1]{\csname bibitem#1\endcsname}%
\let\auto@bib@innerbib\@empty
\bibitem [{\citenamefont {Pei}\ \emph {et~al.}(2011)\citenamefont {Pei},
  \citenamefont {Shi}, \citenamefont {LaLonde}, \citenamefont {Wang},
  \citenamefont {Chen},\ and\ \citenamefont {Snyder}}]{convergince}%
  \BibitemOpen
  \bibfield  {author} {\bibinfo {author} {\bibfnamefont {Y.}~\bibnamefont
  {Pei}}, \bibinfo {author} {\bibfnamefont {X.}~\bibnamefont {Shi}}, \bibinfo
  {author} {\bibfnamefont {A.}~\bibnamefont {LaLonde}}, \bibinfo {author}
  {\bibfnamefont {H.}~\bibnamefont {Wang}}, \bibinfo {author} {\bibfnamefont
  {L.}~\bibnamefont {Chen}}, \ and\ \bibinfo {author} {\bibfnamefont {G.~J.}\
  \bibnamefont {Snyder}},\ }\href {https://doi.org/10.1038/nature09996}
  {\bibfield  {journal} {\bibinfo  {journal} {Nature}\ }\textbf {\bibinfo
  {volume} {473}},\ \bibinfo {pages} {66} (\bibinfo {year} {2011})}\BibitemShut
  {NoStop}%
\bibitem [{\citenamefont {Hicks}\ and\ \citenamefont
  {Dresselhaus}(1993{\natexlab{a}})}]{lowdim}%
  \BibitemOpen
  \bibfield  {author} {\bibinfo {author} {\bibfnamefont {L.~D.}\ \bibnamefont
  {Hicks}}\ and\ \bibinfo {author} {\bibfnamefont {M.~S.}\ \bibnamefont
  {Dresselhaus}},\ }\href@noop {} {\bibfield  {journal} {\bibinfo  {journal}
  {Phys. Rev. B}\ }\textbf {\bibinfo {volume} {47}},\ \bibinfo {pages} {16631}
  (\bibinfo {year} {1993}{\natexlab{a}})}\BibitemShut {NoStop}%
\bibitem [{\citenamefont {Usui}\ and\ \citenamefont {Kuroki}(2017)}]{pudding}%
  \BibitemOpen
  \bibfield  {author} {\bibinfo {author} {\bibfnamefont {H.}~\bibnamefont
  {Usui}}\ and\ \bibinfo {author} {\bibfnamefont {K.}~\bibnamefont {Kuroki}},\
  }\href {\doibase 10.1063/1.4981890} {\bibfield  {journal} {\bibinfo
  {journal} {J. Appl. Phys.}\ }\textbf {\bibinfo {volume} {121}},\ \bibinfo
  {pages} {165101} (\bibinfo {year} {2017})}\BibitemShut {NoStop}%
\bibitem [{\citenamefont {Hicks}\ and\ \citenamefont
  {Dresselhaus}(1993{\natexlab{b}})}]{PhysRevB.47.12727}%
  \BibitemOpen
  \bibfield  {author} {\bibinfo {author} {\bibfnamefont {L.~D.}\ \bibnamefont
  {Hicks}}\ and\ \bibinfo {author} {\bibfnamefont {M.~S.}\ \bibnamefont
  {Dresselhaus}},\ }\href {\doibase 10.1103/PhysRevB.47.12727} {\bibfield
  {journal} {\bibinfo  {journal} {Phys. Rev. B}\ }\textbf {\bibinfo {volume}
  {47}},\ \bibinfo {pages} {12727} (\bibinfo {year}
  {1993}{\natexlab{b}})}\BibitemShut {NoStop}%
\bibitem [{\citenamefont {Yamamoto}\ and\ \citenamefont
  {Fukuyama}(2018)}]{cabon}%
  \BibitemOpen
  \bibfield  {author} {\bibinfo {author} {\bibfnamefont {T.}~\bibnamefont
  {Yamamoto}}\ and\ \bibinfo {author} {\bibfnamefont {H.}~\bibnamefont
  {Fukuyama}},\ }\href {\doibase 10.7566/JPSJ.87.114710} {\bibfield  {journal}
  {\bibinfo  {journal} {J. Phys. Soc. Jpn.}\ }\textbf {\bibinfo {volume}
  {87}},\ \bibinfo {pages} {114710} (\bibinfo {year} {2018})}\BibitemShut
  {NoStop}%
\bibitem [{\citenamefont {Mizuguchi}(2019)}]{JPSJ.88.041001}%
  \BibitemOpen
  \bibfield  {author} {\bibinfo {author} {\bibfnamefont {Y.}~\bibnamefont
  {Mizuguchi}},\ }\href {\doibase 10.7566/JPSJ.88.041001} {\bibfield  {journal}
  {\bibinfo  {journal} {J. Phys. Soc. Jpn.}\ }\textbf {\bibinfo {volume}
  {88}},\ \bibinfo {pages} {041001} (\bibinfo {year} {2019})}\BibitemShut
  {NoStop}%
\bibitem [{\citenamefont {Mizuguchi}\ \emph
  {et~al.}(2012{\natexlab{a}})\citenamefont {Mizuguchi}, \citenamefont
  {Demura}, \citenamefont {Deguchi}, \citenamefont {Takano}, \citenamefont
  {Fujihisa}, \citenamefont {Gotoh}, \citenamefont {Izawa},\ and\ \citenamefont
  {Miura}}]{supercon}%
  \BibitemOpen
  \bibfield  {author} {\bibinfo {author} {\bibfnamefont {Y.}~\bibnamefont
  {Mizuguchi}}, \bibinfo {author} {\bibfnamefont {S.}~\bibnamefont {Demura}},
  \bibinfo {author} {\bibfnamefont {K.}~\bibnamefont {Deguchi}}, \bibinfo
  {author} {\bibfnamefont {Y.}~\bibnamefont {Takano}}, \bibinfo {author}
  {\bibfnamefont {H.}~\bibnamefont {Fujihisa}}, \bibinfo {author}
  {\bibfnamefont {Y.}~\bibnamefont {Gotoh}}, \bibinfo {author} {\bibfnamefont
  {H.}~\bibnamefont {Izawa}}, \ and\ \bibinfo {author} {\bibfnamefont
  {O.}~\bibnamefont {Miura}},\ }\href@noop {} {\bibfield  {journal} {\bibinfo
  {journal} {J. Phys. Soc. Jpn.}\ }\textbf {\bibinfo {volume} {81}},\ \bibinfo
  {pages} {114725} (\bibinfo {year} {2012}{\natexlab{a}})}\BibitemShut
  {NoStop}%
\bibitem [{\citenamefont {Mizuguchi}\ \emph
  {et~al.}(2012{\natexlab{b}})\citenamefont {Mizuguchi}, \citenamefont
  {Fujihisa}, \citenamefont {Gotoh}, \citenamefont {Suzuki}, \citenamefont
  {Usui}, \citenamefont {Kuroki}, \citenamefont {Demura}, \citenamefont
  {Takano}, \citenamefont {Izawa},\ and\ \citenamefont
  {Miura}}]{PhysRevB.86.220510}%
  \BibitemOpen
  \bibfield  {author} {\bibinfo {author} {\bibfnamefont {Y.}~\bibnamefont
  {Mizuguchi}}, \bibinfo {author} {\bibfnamefont {H.}~\bibnamefont {Fujihisa}},
  \bibinfo {author} {\bibfnamefont {Y.}~\bibnamefont {Gotoh}}, \bibinfo
  {author} {\bibfnamefont {K.}~\bibnamefont {Suzuki}}, \bibinfo {author}
  {\bibfnamefont {H.}~\bibnamefont {Usui}}, \bibinfo {author} {\bibfnamefont
  {K.}~\bibnamefont {Kuroki}}, \bibinfo {author} {\bibfnamefont
  {S.}~\bibnamefont {Demura}}, \bibinfo {author} {\bibfnamefont
  {Y.}~\bibnamefont {Takano}}, \bibinfo {author} {\bibfnamefont
  {H.}~\bibnamefont {Izawa}}, \ and\ \bibinfo {author} {\bibfnamefont
  {O.}~\bibnamefont {Miura}},\ }\href {\doibase 10.1103/PhysRevB.86.220510}
  {\bibfield  {journal} {\bibinfo  {journal} {Phys. Rev. B}\ }\textbf {\bibinfo
  {volume} {86}},\ \bibinfo {pages} {220510} (\bibinfo {year}
  {2012}{\natexlab{b}})}\BibitemShut {NoStop}%
\bibitem [{\citenamefont {Mizuguchi}(2015)}]{MIZUGUCHI201534}%
  \BibitemOpen
  \bibfield  {author} {\bibinfo {author} {\bibfnamefont {Y.}~\bibnamefont
  {Mizuguchi}},\ }\href {\doibase https://doi.org/10.1016/j.jpcs.2014.09.003}
  {\bibfield  {journal} {\bibinfo  {journal} {J. Phys. Chem. Solids}\ }\textbf
  {\bibinfo {volume} {84}},\ \bibinfo {pages} {34 } (\bibinfo {year}
  {2015})}\BibitemShut {NoStop}%
\bibitem [{\citenamefont {Yazici}\ \emph {et~al.}(2015)\citenamefont {Yazici},
  \citenamefont {Jeon}, \citenamefont {White},\ and\ \citenamefont
  {Maple}}]{YAZICI2015218}%
  \BibitemOpen
  \bibfield  {author} {\bibinfo {author} {\bibfnamefont {D.}~\bibnamefont
  {Yazici}}, \bibinfo {author} {\bibfnamefont {I.}~\bibnamefont {Jeon}},
  \bibinfo {author} {\bibfnamefont {B.}~\bibnamefont {White}}, \ and\ \bibinfo
  {author} {\bibfnamefont {M.}~\bibnamefont {Maple}},\ }\href {\doibase
  https://doi.org/10.1016/j.physc.2015.02.025} {\bibfield  {journal} {\bibinfo
  {journal} {Physica C: Superconductivity and its Applications}\ }\textbf
  {\bibinfo {volume} {514}},\ \bibinfo {pages} {218 } (\bibinfo {year}
  {2015})}\BibitemShut {NoStop}%
\bibitem [{\citenamefont {Usui}\ and\ \citenamefont
  {Kuroki}(2015)}]{supcomd.mat1.50.2015}%
  \BibitemOpen
  \bibfield  {author} {\bibinfo {author} {\bibfnamefont {H.}~\bibnamefont
  {Usui}}\ and\ \bibinfo {author} {\bibfnamefont {K.}~\bibnamefont {Kuroki}},\
  }\href@noop {} {\bibfield  {journal} {\bibinfo  {journal} {Novel. Supercond.
  Mater.}\ }\textbf {\bibinfo {volume} {1}},\ \bibinfo {pages} {50} (\bibinfo
  {year} {2015})}\BibitemShut {NoStop}%
\bibitem [{\citenamefont {Mizuguchi}\ \emph {et~al.}(2016)\citenamefont
  {Mizuguchi}, \citenamefont {Nishida}, \citenamefont {Omachi},\ and\
  \citenamefont {Miura}}]{series}%
  \BibitemOpen
  \bibfield  {author} {\bibinfo {author} {\bibfnamefont {Y.}~\bibnamefont
  {Mizuguchi}}, \bibinfo {author} {\bibfnamefont {A.}~\bibnamefont {Nishida}},
  \bibinfo {author} {\bibfnamefont {A.}~\bibnamefont {Omachi}}, \ and\ \bibinfo
  {author} {\bibfnamefont {O.}~\bibnamefont {Miura}},\ }\href
  {http://doi.org/10.1080/23311940.2016.1156281} {\bibfield  {journal}
  {\bibinfo  {journal} {Cogent Physics}\ }\textbf {\bibinfo {volume} {3}}
  (\bibinfo {year} {2016})}\BibitemShut {NoStop}%
\bibitem [{\citenamefont {Usui}\ \emph {et~al.}(2012)\citenamefont {Usui},
  \citenamefont {Suzuki},\ and\ \citenamefont {Kuroki}}]{quasiband}%
  \BibitemOpen
  \bibfield  {author} {\bibinfo {author} {\bibfnamefont {H.}~\bibnamefont
  {Usui}}, \bibinfo {author} {\bibfnamefont {K.}~\bibnamefont {Suzuki}}, \ and\
  \bibinfo {author} {\bibfnamefont {K.}~\bibnamefont {Kuroki}},\ }\href@noop {}
  {\bibfield  {journal} {\bibinfo  {journal} {Phys. Rev. B}\ }\textbf {\bibinfo
  {volume} {86}},\ \bibinfo {pages} {220501(R)} (\bibinfo {year}
  {2012})}\BibitemShut {NoStop}%
\bibitem [{\citenamefont {Ochi}\ \emph {et~al.}(2017)\citenamefont {Ochi},
  \citenamefont {Usui},\ and\ \citenamefont {Kuroki}}]{LaOAsSe2}%
  \BibitemOpen
  \bibfield  {author} {\bibinfo {author} {\bibfnamefont {M.}~\bibnamefont
  {Ochi}}, \bibinfo {author} {\bibfnamefont {H.}~\bibnamefont {Usui}}, \ and\
  \bibinfo {author} {\bibfnamefont {K.}~\bibnamefont {Kuroki}},\ }\href@noop {}
  {\bibfield  {journal} {\bibinfo  {journal} {Phys. Rev. Applied}\ }\textbf
  {\bibinfo {volume} {8}},\ \bibinfo {pages} {064020} (\bibinfo {year}
  {2017})}\BibitemShut {NoStop}%
\bibitem [{\citenamefont {Ochi}\ \emph {et~al.}(2019)\citenamefont {Ochi},
  \citenamefont {Usui},\ and\ \citenamefont {Kuroki}}]{JPSJ.88.041010}%
  \BibitemOpen
  \bibfield  {author} {\bibinfo {author} {\bibfnamefont {M.}~\bibnamefont
  {Ochi}}, \bibinfo {author} {\bibfnamefont {H.}~\bibnamefont {Usui}}, \ and\
  \bibinfo {author} {\bibfnamefont {K.}~\bibnamefont {Kuroki}},\ }\href
  {\doibase 10.7566/JPSJ.88.041010} {\bibfield  {journal} {\bibinfo  {journal}
  {J. Phys. Soc. Jpn.}\ }\textbf {\bibinfo {volume} {88}},\ \bibinfo {pages}
  {041010} (\bibinfo {year} {2019})}\BibitemShut {NoStop}%
\bibitem [{\citenamefont {Lee}\ \emph {et~al.}(2018)\citenamefont {Lee},
  \citenamefont {Nishida}, \citenamefont {Hasegawa}, \citenamefont {Nishiate},
  \citenamefont {Kunioka}, \citenamefont {Ohira-Kawamura}, \citenamefont
  {Nakamura}, \citenamefont {Nakajima},\ and\ \citenamefont
  {Mizuguchi}}]{lattring}%
  \BibitemOpen
  \bibfield  {author} {\bibinfo {author} {\bibfnamefont {C.~H.}\ \bibnamefont
  {Lee}}, \bibinfo {author} {\bibfnamefont {A.}~\bibnamefont {Nishida}},
  \bibinfo {author} {\bibfnamefont {T.}~\bibnamefont {Hasegawa}}, \bibinfo
  {author} {\bibfnamefont {H.}~\bibnamefont {Nishiate}}, \bibinfo {author}
  {\bibfnamefont {H.}~\bibnamefont {Kunioka}}, \bibinfo {author} {\bibfnamefont
  {S.}~\bibnamefont {Ohira-Kawamura}}, \bibinfo {author} {\bibfnamefont
  {M.}~\bibnamefont {Nakamura}}, \bibinfo {author} {\bibfnamefont
  {K.}~\bibnamefont {Nakajima}}, \ and\ \bibinfo {author} {\bibfnamefont
  {Y.}~\bibnamefont {Mizuguchi}},\ }\href {\doibase 10.1063/1.5010373}
  {\bibfield  {journal} {\bibinfo  {journal} {Appl. Phys. Lett.}\ }\textbf
  {\bibinfo {volume} {112}},\ \bibinfo {pages} {023903} (\bibinfo {year}
  {2018})}\BibitemShut {NoStop}%
\bibitem [{\citenamefont {Lee}(2019)}]{JPSJ.88.041009}%
  \BibitemOpen
  \bibfield  {author} {\bibinfo {author} {\bibfnamefont {C.-H.}\ \bibnamefont
  {Lee}},\ }\href {\doibase 10.7566/JPSJ.88.041009} {\bibfield  {journal}
  {\bibinfo  {journal} {J. Phys. Soc. Jpn.}\ }\textbf {\bibinfo {volume}
  {88}},\ \bibinfo {pages} {041009} (\bibinfo {year} {2019})}\BibitemShut
  {NoStop}%
\bibitem [{\citenamefont {Nishida}\ \emph {et~al.}(2015)\citenamefont
  {Nishida}, \citenamefont {Miura}, \citenamefont {Lee},\ and\ \citenamefont
  {Mizuguchi}}]{SSeZT}%
  \BibitemOpen
  \bibfield  {author} {\bibinfo {author} {\bibfnamefont {A.}~\bibnamefont
  {Nishida}}, \bibinfo {author} {\bibfnamefont {O.}~\bibnamefont {Miura}},
  \bibinfo {author} {\bibfnamefont {C.-H.}\ \bibnamefont {Lee}}, \ and\
  \bibinfo {author} {\bibfnamefont {Y.}~\bibnamefont {Mizuguchi}},\ }\href@noop
  {} {\bibfield  {journal} {\bibinfo  {journal} {Appl. Phys. Express}\ }\textbf
  {\bibinfo {volume} {8}},\ \bibinfo {pages} {111801} (\bibinfo {year}
  {2015})}\BibitemShut {NoStop}%
\bibitem [{\citenamefont {Goto}\ \emph {et~al.}(2018)\citenamefont {Goto},
  \citenamefont {Miura}, \citenamefont {Sakagami}, \citenamefont {Kamihara},
  \citenamefont {Moriyoshi}, \citenamefont {Kuroiwa},\ and\ \citenamefont
  {Mizuguchi}}]{JPSJ.87.074703}%
  \BibitemOpen
  \bibfield  {author} {\bibinfo {author} {\bibfnamefont {Y.}~\bibnamefont
  {Goto}}, \bibinfo {author} {\bibfnamefont {A.}~\bibnamefont {Miura}},
  \bibinfo {author} {\bibfnamefont {R.}~\bibnamefont {Sakagami}}, \bibinfo
  {author} {\bibfnamefont {Y.}~\bibnamefont {Kamihara}}, \bibinfo {author}
  {\bibfnamefont {C.}~\bibnamefont {Moriyoshi}}, \bibinfo {author}
  {\bibfnamefont {Y.}~\bibnamefont {Kuroiwa}}, \ and\ \bibinfo {author}
  {\bibfnamefont {Y.}~\bibnamefont {Mizuguchi}},\ }\href {\doibase
  10.7566/JPSJ.87.074703} {\bibfield  {journal} {\bibinfo  {journal} {J. Phys.
  Soc. Jpn.}\ }\textbf {\bibinfo {volume} {87}},\ \bibinfo {pages} {074703}
  (\bibinfo {year} {2018})}\BibitemShut {NoStop}%
\bibitem [{\citenamefont {Goto}\ \emph {et~al.}(2019)\citenamefont {Goto},
  \citenamefont {Miura}, \citenamefont {Moriyoshi}, \citenamefont {Kuroiwa},\
  and\ \citenamefont {Mizuguchi}}]{JPSJ880247052019}%
  \BibitemOpen
  \bibfield  {author} {\bibinfo {author} {\bibfnamefont {Y.}~\bibnamefont
  {Goto}}, \bibinfo {author} {\bibfnamefont {A.}~\bibnamefont {Miura}},
  \bibinfo {author} {\bibfnamefont {C.}~\bibnamefont {Moriyoshi}}, \bibinfo
  {author} {\bibfnamefont {Y.}~\bibnamefont {Kuroiwa}}, \ and\ \bibinfo
  {author} {\bibfnamefont {Y.}~\bibnamefont {Mizuguchi}},\ }\href {\doibase
  10.7566/JPSJ.88.024705} {\bibfield  {journal} {\bibinfo  {journal} {J. Phys.
  Soc. Jpn.}\ }\textbf {\bibinfo {volume} {88}},\ \bibinfo {pages} {024705}
  (\bibinfo {year} {2019})}\BibitemShut {NoStop}%
\bibitem [{\citenamefont {Sun}\ \emph {et~al.}(2014)\citenamefont {Sun},
  \citenamefont {Ablimit}, \citenamefont {Zhai}, \citenamefont {Bao},
  \citenamefont {Tang}, \citenamefont {Wang}, \citenamefont {Wang},
  \citenamefont {Feng},\ and\ \citenamefont {Cao}}]{firstsynth}%
  \BibitemOpen
  \bibfield  {author} {\bibinfo {author} {\bibfnamefont {Y.-L.}\ \bibnamefont
  {Sun}}, \bibinfo {author} {\bibfnamefont {A.}~\bibnamefont {Ablimit}},
  \bibinfo {author} {\bibfnamefont {H.-F.}\ \bibnamefont {Zhai}}, \bibinfo
  {author} {\bibfnamefont {J.-K.}\ \bibnamefont {Bao}}, \bibinfo {author}
  {\bibfnamefont {Z.-T.}\ \bibnamefont {Tang}}, \bibinfo {author}
  {\bibfnamefont {X.-B.}\ \bibnamefont {Wang}}, \bibinfo {author}
  {\bibfnamefont {N.-L.}\ \bibnamefont {Wang}}, \bibinfo {author}
  {\bibfnamefont {C.-M.}\ \bibnamefont {Feng}}, \ and\ \bibinfo {author}
  {\bibfnamefont {G.-H.}\ \bibnamefont {Cao}},\ }\href {\doibase
  10.1021/ic501687h} {\bibfield  {journal} {\bibinfo  {journal} {Inorg. Chem.}\
  }\textbf {\bibinfo {volume} {53}},\ \bibinfo {pages} {11125} (\bibinfo {year}
  {2014})}\BibitemShut {NoStop}%
\bibitem [{\citenamefont {Mizuguchi}\ \emph {et~al.}(2017)\citenamefont
  {Mizuguchi}, \citenamefont {Hijikata}, \citenamefont {Abe}, \citenamefont
  {Moriyoshi}, \citenamefont {Kuroiwa}, \citenamefont {Goto}, \citenamefont
  {Miura}, \citenamefont {Lee}, \citenamefont {Torii}, \citenamefont
  {Kamiyama}, \citenamefont {Lee}, \citenamefont {Ochi},\ and\ \citenamefont
  {Kuroki}}]{siteslect}%
  \BibitemOpen
  \bibfield  {author} {\bibinfo {author} {\bibfnamefont {Y.}~\bibnamefont
  {Mizuguchi}}, \bibinfo {author} {\bibfnamefont {Y.}~\bibnamefont {Hijikata}},
  \bibinfo {author} {\bibfnamefont {T.}~\bibnamefont {Abe}}, \bibinfo {author}
  {\bibfnamefont {C.}~\bibnamefont {Moriyoshi}}, \bibinfo {author}
  {\bibfnamefont {Y.}~\bibnamefont {Kuroiwa}}, \bibinfo {author} {\bibfnamefont
  {Y.}~\bibnamefont {Goto}}, \bibinfo {author} {\bibfnamefont {A.}~\bibnamefont
  {Miura}}, \bibinfo {author} {\bibfnamefont {S.}~\bibnamefont {Lee}}, \bibinfo
  {author} {\bibfnamefont {S.}~\bibnamefont {Torii}}, \bibinfo {author}
  {\bibfnamefont {T.}~\bibnamefont {Kamiyama}}, \bibinfo {author}
  {\bibfnamefont {C.~H.}\ \bibnamefont {Lee}}, \bibinfo {author} {\bibfnamefont
  {M.}~\bibnamefont {Ochi}}, \ and\ \bibinfo {author} {\bibfnamefont
  {K.}~\bibnamefont {Kuroki}},\ }\href
  {http://stacks.iop.org/0295-5075/119/i=2/a=26002} {\bibfield  {journal}
  {\bibinfo  {journal} {EPL}\ }\textbf {\bibinfo {volume} {119}},\ \bibinfo
  {pages} {26002} (\bibinfo {year} {2017})}\BibitemShut {NoStop}%
\bibitem [{\citenamefont {Yu}\ \emph {et~al.}(2019)\citenamefont {Yu},
  \citenamefont {Wang}, \citenamefont {Li}, \citenamefont {Cheng},
  \citenamefont {Wang},\ and\ \citenamefont {Zhang}}]{fdope}%
  \BibitemOpen
  \bibfield  {author} {\bibinfo {author} {\bibfnamefont {Y.}~\bibnamefont
  {Yu}}, \bibinfo {author} {\bibfnamefont {C.}~\bibnamefont {Wang}}, \bibinfo
  {author} {\bibfnamefont {Q.}~\bibnamefont {Li}}, \bibinfo {author}
  {\bibfnamefont {C.}~\bibnamefont {Cheng}}, \bibinfo {author} {\bibfnamefont
  {S.}~\bibnamefont {Wang}}, \ and\ \bibinfo {author} {\bibfnamefont
  {C.}~\bibnamefont {Zhang}},\ }\href {\doibase
  https://doi.org/10.1016/j.ceramint.2018.09.248} {\bibfield  {journal}
  {\bibinfo  {journal} {Ceram. Int.}\ }\textbf {\bibinfo {volume} {45}},\
  \bibinfo {pages} {817 } (\bibinfo {year} {2019})}\BibitemShut {NoStop}%
\bibitem [{\citenamefont {Momma}\ and\ \citenamefont {Izumi}(2011)}]{VESTA}%
  \BibitemOpen
  \bibfield  {author} {\bibinfo {author} {\bibfnamefont {K.}~\bibnamefont
  {Momma}}\ and\ \bibinfo {author} {\bibfnamefont {F.}~\bibnamefont {Izumi}},\
  }\href {\doibase 10.1107/S0021889811038970} {\bibfield  {journal} {\bibinfo
  {journal} {J. Appl. Crystallogr}\ }\textbf {\bibinfo {volume} {44}},\
  \bibinfo {pages} {1272} (\bibinfo {year} {2011})}\BibitemShut {NoStop}%
\bibitem [{\citenamefont {Perdew}\ \emph {et~al.}(2008)\citenamefont {Perdew},
  \citenamefont {Ruzsinszky}, \citenamefont {Csonka}, \citenamefont {Vydrov},
  \citenamefont {Scuseria}, \citenamefont {Constantin}, \citenamefont {Zhou},\
  and\ \citenamefont {Burke}}]{PBEsol}%
  \BibitemOpen
  \bibfield  {author} {\bibinfo {author} {\bibfnamefont {J.~P.}\ \bibnamefont
  {Perdew}}, \bibinfo {author} {\bibfnamefont {A.}~\bibnamefont {Ruzsinszky}},
  \bibinfo {author} {\bibfnamefont {G.~I.}\ \bibnamefont {Csonka}}, \bibinfo
  {author} {\bibfnamefont {O.~A.}\ \bibnamefont {Vydrov}}, \bibinfo {author}
  {\bibfnamefont {G.~E.}\ \bibnamefont {Scuseria}}, \bibinfo {author}
  {\bibfnamefont {L.~A.}\ \bibnamefont {Constantin}}, \bibinfo {author}
  {\bibfnamefont {X.}~\bibnamefont {Zhou}}, \ and\ \bibinfo {author}
  {\bibfnamefont {K.}~\bibnamefont {Burke}},\ }\href@noop {} {\bibfield
  {journal} {\bibinfo  {journal} {Phys. Rev. Lett.}\ }\textbf {\bibinfo
  {volume} {100}},\ \bibinfo {pages} {136406} (\bibinfo {year}
  {2008})}\BibitemShut {NoStop}%
\bibitem [{\citenamefont {Kresse}\ and\ \citenamefont
  {Joubert}(1999)}]{PhysRevB.59.1758}%
  \BibitemOpen
  \bibfield  {author} {\bibinfo {author} {\bibfnamefont {G.}~\bibnamefont
  {Kresse}}\ and\ \bibinfo {author} {\bibfnamefont {D.}~\bibnamefont
  {Joubert}},\ }\href {\doibase 10.1103/PhysRevB.59.1758} {\bibfield  {journal}
  {\bibinfo  {journal} {Phys. Rev. B}\ }\textbf {\bibinfo {volume} {59}},\
  \bibinfo {pages} {1758} (\bibinfo {year} {1999})}\BibitemShut {NoStop}%
\bibitem [{\citenamefont {Kresse}\ and\ \citenamefont {Hafner}(1993)}]{vasp1}%
  \BibitemOpen
  \bibfield  {author} {\bibinfo {author} {\bibfnamefont {G.}~\bibnamefont
  {Kresse}}\ and\ \bibinfo {author} {\bibfnamefont {J.}~\bibnamefont
  {Hafner}},\ }\href@noop {} {\bibfield  {journal} {\bibinfo  {journal} {Phys.
  Rev. B}\ }\textbf {\bibinfo {volume} {47}},\ \bibinfo {pages} {559} (\bibinfo
  {year} {1993})}\BibitemShut {NoStop}%
\bibitem [{\citenamefont {Kresse}\ and\ \citenamefont {Hafner}(1994)}]{vasp2}%
  \BibitemOpen
  \bibfield  {author} {\bibinfo {author} {\bibfnamefont {G.}~\bibnamefont
  {Kresse}}\ and\ \bibinfo {author} {\bibfnamefont {J.}~\bibnamefont
  {Hafner}},\ }\href@noop {} {\bibfield  {journal} {\bibinfo  {journal} {Phys.
  Rev. B}\ }\textbf {\bibinfo {volume} {49}},\ \bibinfo {pages} {14251}
  (\bibinfo {year} {1994})}\BibitemShut {NoStop}%
\bibitem [{\citenamefont {Kresse}\ and\ \citenamefont
  {Furthm{\"u}ller}(1996)}]{vasp3}%
  \BibitemOpen
  \bibfield  {author} {\bibinfo {author} {\bibfnamefont {G.}~\bibnamefont
  {Kresse}}\ and\ \bibinfo {author} {\bibfnamefont {J.}~\bibnamefont
  {Furthm{\"u}ller}},\ }\href {\doibase
  https://doi.org/10.1016/0927-0256(96)00008-0} {\bibfield  {journal} {\bibinfo
   {journal} {Comput. Mater. Sci.}\ }\textbf {\bibinfo {volume} {6}},\ \bibinfo
  {pages} {15 } (\bibinfo {year} {1996})}\BibitemShut {NoStop}%
\bibitem [{\citenamefont {Kresse}\ and\ \citenamefont
  {Furthm\"uller}(1996)}]{vasp4}%
  \BibitemOpen
  \bibfield  {author} {\bibinfo {author} {\bibfnamefont {G.}~\bibnamefont
  {Kresse}}\ and\ \bibinfo {author} {\bibfnamefont {J.}~\bibnamefont
  {Furthm\"uller}},\ }\href@noop {} {\bibfield  {journal} {\bibinfo  {journal}
  {Phys. Rev. B}\ }\textbf {\bibinfo {volume} {54}},\ \bibinfo {pages} {11169}
  (\bibinfo {year} {1996})}\BibitemShut {NoStop}%
\bibitem [{\citenamefont {Becke}\ and\ \citenamefont {Johnson}(2006)}]{mbj1}%
  \BibitemOpen
  \bibfield  {author} {\bibinfo {author} {\bibfnamefont {A.~D.}\ \bibnamefont
  {Becke}}\ and\ \bibinfo {author} {\bibfnamefont {E.~R.}\ \bibnamefont
  {Johnson}},\ }\href {\doibase 10.1063/1.2213970} {\bibfield  {journal}
  {\bibinfo  {journal} {J. Chem. Phys.}\ }\textbf {\bibinfo {volume} {124}},\
  \bibinfo {pages} {221101} (\bibinfo {year} {2006})}\BibitemShut {NoStop}%
\bibitem [{\citenamefont {Tran}\ and\ \citenamefont {Blaha}(2009)}]{mbj2}%
  \BibitemOpen
  \bibfield  {author} {\bibinfo {author} {\bibfnamefont {F.}~\bibnamefont
  {Tran}}\ and\ \bibinfo {author} {\bibfnamefont {P.}~\bibnamefont {Blaha}},\
  }\href@noop {} {\bibfield  {journal} {\bibinfo  {journal} {Phys. Rev. Lett.}\
  }\textbf {\bibinfo {volume} {102}},\ \bibinfo {pages} {226401} (\bibinfo
  {year} {2009})}\BibitemShut {NoStop}%
\bibitem [{\citenamefont {Blaha}\ \emph {et~al.}(2001)\citenamefont {Blaha},
  \citenamefont {Schwarz}, \citenamefont {Madsen}, \citenamefont {Kvasnicka},\
  and\ \citenamefont {Luitz}}]{wien2k}%
  \BibitemOpen
  \bibfield  {author} {\bibinfo {author} {\bibfnamefont {P.}~\bibnamefont
  {Blaha}}, \bibinfo {author} {\bibfnamefont {K.}~\bibnamefont {Schwarz}},
  \bibinfo {author} {\bibfnamefont {G.~K.~H.}\ \bibnamefont {Madsen}}, \bibinfo
  {author} {\bibfnamefont {D.}~\bibnamefont {Kvasnicka}}, \ and\ \bibinfo
  {author} {\bibfnamefont {J.}~\bibnamefont {Luitz}},\ }\href@noop {} {\emph
  {\bibinfo {title} {WIEN2K: An Augmented Plane Wave Plus Local Orbitals
  Program for Calculating Crystal Properties}}},\ \bibinfo {organization}
  {Technische Universit{\"a}t Wien, Vienna,} (\bibinfo {year}
  {2001})\BibitemShut {NoStop}%
\bibitem [{\citenamefont {Madsen}\ and\ \citenamefont
  {Singh}(2006)}]{BoltzTraP}%
  \BibitemOpen
  \bibfield  {author} {\bibinfo {author} {\bibfnamefont {G.}~\bibnamefont
  {Madsen}}\ and\ \bibinfo {author} {\bibfnamefont {D.}~\bibnamefont {Singh}},\
  }\href@noop {} {\bibfield  {journal} {\bibinfo  {journal} {Comput. Phys.
  Commun.}\ }\textbf {\bibinfo {volume} {175}} (\bibinfo {year}
  {2006})}\BibitemShut {NoStop}%
\bibitem [{\citenamefont {Mostofi}\ \emph {et~al.}(2008)\citenamefont
  {Mostofi}, \citenamefont {Yates}, \citenamefont {Lee}, \citenamefont {Souza},
  \citenamefont {Vanderbilt},\ and\ \citenamefont {Marzari}}]{MOSTOFI2008685}%
  \BibitemOpen
  \bibfield  {author} {\bibinfo {author} {\bibfnamefont {A.~A.}\ \bibnamefont
  {Mostofi}}, \bibinfo {author} {\bibfnamefont {J.~R.}\ \bibnamefont {Yates}},
  \bibinfo {author} {\bibfnamefont {Y.-S.}\ \bibnamefont {Lee}}, \bibinfo
  {author} {\bibfnamefont {I.}~\bibnamefont {Souza}}, \bibinfo {author}
  {\bibfnamefont {D.}~\bibnamefont {Vanderbilt}}, \ and\ \bibinfo {author}
  {\bibfnamefont {N.}~\bibnamefont {Marzari}},\ }\href {\doibase
  https://doi.org/10.1016/j.cpc.2007.11.016} {\bibfield  {journal} {\bibinfo
  {journal} {Comput. Phys. Commun.}\ }\textbf {\bibinfo {volume} {178}},\
  \bibinfo {pages} {685 } (\bibinfo {year} {2008})}\BibitemShut {NoStop}%
\bibitem [{\citenamefont {Kune{\v s}}\ \emph {et~al.}(2010)\citenamefont
  {Kune{\v s}}, \citenamefont {Arita}, \citenamefont {Wissgott}, \citenamefont
  {Toschi}, \citenamefont {Ikeda},\ and\ \citenamefont {Held}}]{KUNES20101888}%
  \BibitemOpen
  \bibfield  {author} {\bibinfo {author} {\bibfnamefont {J.}~\bibnamefont
  {Kune{\v s}}}, \bibinfo {author} {\bibfnamefont {R.}~\bibnamefont {Arita}},
  \bibinfo {author} {\bibfnamefont {P.}~\bibnamefont {Wissgott}}, \bibinfo
  {author} {\bibfnamefont {A.}~\bibnamefont {Toschi}}, \bibinfo {author}
  {\bibfnamefont {H.}~\bibnamefont {Ikeda}}, \ and\ \bibinfo {author}
  {\bibfnamefont {K.}~\bibnamefont {Held}},\ }\href {\doibase
  https://doi.org/10.1016/j.cpc.2010.08.005} {\bibfield  {journal} {\bibinfo
  {journal} {Comput. Phys. Commun.}\ }\textbf {\bibinfo {volume} {181}},\
  \bibinfo {pages} {1888 } (\bibinfo {year} {2010})}\BibitemShut {NoStop}%
\bibitem [{\citenamefont {Souza}\ \emph {et~al.}(2001)\citenamefont {Souza},
  \citenamefont {Marzari},\ and\ \citenamefont
  {Vanderbilt}}]{PhysRevB.65.035109}%
  \BibitemOpen
  \bibfield  {author} {\bibinfo {author} {\bibfnamefont {I.}~\bibnamefont
  {Souza}}, \bibinfo {author} {\bibfnamefont {N.}~\bibnamefont {Marzari}}, \
  and\ \bibinfo {author} {\bibfnamefont {D.}~\bibnamefont {Vanderbilt}},\
  }\href {\doibase 10.1103/PhysRevB.65.035109} {\bibfield  {journal} {\bibinfo
  {journal} {Phys. Rev. B}\ }\textbf {\bibinfo {volume} {65}},\ \bibinfo
  {pages} {035109} (\bibinfo {year} {2001})}\BibitemShut {NoStop}%
\bibitem [{\citenamefont {Marzari}\ and\ \citenamefont
  {Vanderbilt}(1997)}]{PhysRevB.56.12847}%
  \BibitemOpen
  \bibfield  {author} {\bibinfo {author} {\bibfnamefont {N.}~\bibnamefont
  {Marzari}}\ and\ \bibinfo {author} {\bibfnamefont {D.}~\bibnamefont
  {Vanderbilt}},\ }\href {\doibase 10.1103/PhysRevB.56.12847} {\bibfield
  {journal} {\bibinfo  {journal} {Phys. Rev. B}\ }\textbf {\bibinfo {volume}
  {56}},\ \bibinfo {pages} {12847} (\bibinfo {year} {1997})}\BibitemShut
  {NoStop}%
\bibitem [{\citenamefont {Ochi}\ \emph {et~al.}(2016)\citenamefont {Ochi},
  \citenamefont {Akashi},\ and\ \citenamefont {Kuroki}}]{JPSJ.85.094705}%
  \BibitemOpen
  \bibfield  {author} {\bibinfo {author} {\bibfnamefont {M.}~\bibnamefont
  {Ochi}}, \bibinfo {author} {\bibfnamefont {R.}~\bibnamefont {Akashi}}, \ and\
  \bibinfo {author} {\bibfnamefont {K.}~\bibnamefont {Kuroki}},\ }\href
  {\doibase 10.7566/JPSJ.85.094705} {\bibfield  {journal} {\bibinfo  {journal}
  {J. Phys. Soc. Jpn.}\ }\textbf {\bibinfo {volume} {85}},\ \bibinfo {pages}
  {094705} (\bibinfo {year} {2016})}\BibitemShut {NoStop}%
\bibitem [{\citenamefont {Kuroki}\ and\ \citenamefont
  {Arita}(2007)}]{JPSJ.76.083707}%
  \BibitemOpen
  \bibfield  {author} {\bibinfo {author} {\bibfnamefont {K.}~\bibnamefont
  {Kuroki}}\ and\ \bibinfo {author} {\bibfnamefont {R.}~\bibnamefont {Arita}},\
  }\href {\doibase 10.1143/JPSJ.76.083707} {\bibfield  {journal} {\bibinfo
  {journal} {J. Phys. Soc. Jpn.}\ }\textbf {\bibinfo {volume} {76}},\ \bibinfo
  {pages} {083707} (\bibinfo {year} {2007})}\BibitemShut {NoStop}%
\bibitem [{\citenamefont {Usui}\ \emph {et~al.}(2009)\citenamefont {Usui},
  \citenamefont {Arita},\ and\ \citenamefont {Kuroki}}]{Usui_2009}%
  \BibitemOpen
  \bibfield  {author} {\bibinfo {author} {\bibfnamefont {H.}~\bibnamefont
  {Usui}}, \bibinfo {author} {\bibfnamefont {R.}~\bibnamefont {Arita}}, \ and\
  \bibinfo {author} {\bibfnamefont {K.}~\bibnamefont {Kuroki}},\ }\href
  {\doibase 10.1088/0953-8984/21/6/064223} {\bibfield  {journal} {\bibinfo
  {journal} {J. Phys. Condens. Matter}\ }\textbf {\bibinfo {volume} {21}},\
  \bibinfo {pages} {064223} (\bibinfo {year} {2009})}\BibitemShut {NoStop}%
\bibitem [{\citenamefont {Usui}\ \emph {et~al.}(2013)\citenamefont {Usui},
  \citenamefont {Suzuki}, \citenamefont {Kuroki}, \citenamefont {Nakano},
  \citenamefont {Kudo},\ and\ \citenamefont {Nohara}}]{PhysRevB.88.075140}%
  \BibitemOpen
  \bibfield  {author} {\bibinfo {author} {\bibfnamefont {H.}~\bibnamefont
  {Usui}}, \bibinfo {author} {\bibfnamefont {K.}~\bibnamefont {Suzuki}},
  \bibinfo {author} {\bibfnamefont {K.}~\bibnamefont {Kuroki}}, \bibinfo
  {author} {\bibfnamefont {S.}~\bibnamefont {Nakano}}, \bibinfo {author}
  {\bibfnamefont {K.}~\bibnamefont {Kudo}}, \ and\ \bibinfo {author}
  {\bibfnamefont {M.}~\bibnamefont {Nohara}},\ }\href {\doibase
  10.1103/PhysRevB.88.075140} {\bibfield  {journal} {\bibinfo  {journal} {Phys.
  Rev. B}\ }\textbf {\bibinfo {volume} {88}},\ \bibinfo {pages} {075140}
  (\bibinfo {year} {2013})}\BibitemShut {NoStop}%
\bibitem [{\citenamefont {Mori}\ \emph {et~al.}(2013)\citenamefont {Mori},
  \citenamefont {Sakakibara}, \citenamefont {Usui},\ and\ \citenamefont
  {Kuroki}}]{PhysRevB.88.075141}%
  \BibitemOpen
  \bibfield  {author} {\bibinfo {author} {\bibfnamefont {K.}~\bibnamefont
  {Mori}}, \bibinfo {author} {\bibfnamefont {H.}~\bibnamefont {Sakakibara}},
  \bibinfo {author} {\bibfnamefont {H.}~\bibnamefont {Usui}}, \ and\ \bibinfo
  {author} {\bibfnamefont {K.}~\bibnamefont {Kuroki}},\ }\href {\doibase
  10.1103/PhysRevB.88.075141} {\bibfield  {journal} {\bibinfo  {journal} {Phys.
  Rev. B}\ }\textbf {\bibinfo {volume} {88}},\ \bibinfo {pages} {075141}
  (\bibinfo {year} {2013})}\BibitemShut {NoStop}%
\bibitem [{\citenamefont {Usui}\ \emph {et~al.}(2014)\citenamefont {Usui},
  \citenamefont {Kuroki}, \citenamefont {Nakano}, \citenamefont {Kudo},\ and\
  \citenamefont {Nohara}}]{Usui2014}%
  \BibitemOpen
  \bibfield  {author} {\bibinfo {author} {\bibfnamefont {H.}~\bibnamefont
  {Usui}}, \bibinfo {author} {\bibfnamefont {K.}~\bibnamefont {Kuroki}},
  \bibinfo {author} {\bibfnamefont {S.}~\bibnamefont {Nakano}}, \bibinfo
  {author} {\bibfnamefont {K.}~\bibnamefont {Kudo}}, \ and\ \bibinfo {author}
  {\bibfnamefont {M.}~\bibnamefont {Nohara}},\ }\href {\doibase
  10.1007/s11664-013-2823-5} {\bibfield  {journal} {\bibinfo  {journal} {J.
  Electron. Mater}\ }\textbf {\bibinfo {volume} {43}},\ \bibinfo {pages} {1656}
  (\bibinfo {year} {2014})}\BibitemShut {NoStop}%
\end{thebibliography}%
\end{document}